\documentclass[11pt,english]{article}

\usepackage[T1]{fontenc}
\usepackage[utf8]{inputenc}
\usepackage{lmodern}
\usepackage{amsmath}
\usepackage{amssymb}
\usepackage{mathrsfs}
\usepackage{cancel}
\usepackage[square,numbers,sort&compress]{natbib}
\usepackage[english]{babel}
\usepackage{physics}

\usepackage{geometry}
\usepackage{caption}
\usepackage{subcaption}
\usepackage{graphicx}
\graphicspath{{Figures/}}
\usepackage[toc,page]{appendix}

\usepackage{epstopdf} 
\usepackage{comment}
\usepackage{tabularx} 
\usepackage{booktabs} 
\usepackage{psfrag} 

\usepackage{xcolor} 
\usepackage{empheq} 

\definecolor{linkcolor}{rgb}{0,0,0.6} 
\usepackage[ 
    pdftex,
    colorlinks=true,
    pdfstartview=FitV,
    linkcolor=linkcolor,
    citecolor=linkcolor,
    urlcolor=linkcolor,
    hyperindex=true,
    hyperfigures=false,
    backref=page,
]{hyperref}
\renewcommand*{\backref}[1]{}
\renewcommand*{\backrefalt}[4]{%
  \ifcase #1 
  \or
    (cit. on p.~#2)%
  \else
    (cit. on pp.~#2)%
  \fi
}

\usepackage{fancyhdr} 
\usepackage{pgfplots} 
\usepgfplotslibrary{groupplots}

\newcommand\mv{\sigma}
\newcommand\vv{\mathscr{V}}
\newcommand\zv{\mathscr{Z}}
\newcommand\nv{\mathscr{N}}
\newcommand\Ne{\mathscr{E}_1}
\newcommand\alg{\mathscr{E}_2}

\newcommand\rmax{r_*^{\max}}
\newcommand\rh{r_{h}}
\newcommand\phiback{\bar{\phi}}
\newcommand\gmunuback{\bar{g}_{\mu \nu}}

\newcommand\Veff{V_{\mathrm{eff}}}

\newcommand\mreg{M_p^{\mathrm{reg}}} 
\newcommand\Bterm{\mathscr{B}}
\newcommand\Dope{\mathscr{D}}
\newcommand\Schroop{{\mathscr{L}}}
\newcommand\SchroopF{\Schroop_{\mathrm{F}}}
\newcommand\SchroopEt{\Schroop_{\theta}}

\newcommand\cA{\mathcal{A}}
\newcommand\cB{\mathcal{B}}

\newcommand\nth{{p}}
\newcommand\Ftwo{{F}}

\newcommand\Det{\Delta}

\newcommand\reff{r_0}
\newcommand\rsc{r_m}
\newcommand\Vs{V_{S}}
\newcommand\smv{\zeta}
\newcommand\zeromode{\smv_0}
\newcommand\zeromodeMono{\Psi_0}
\newcommand\smono{S_{\ell=0}}
\newcommand\sdipo{S_{\ell=1}}

\providecommand{\GT}{\mathcal{G}_T}
\providecommand{\FT}{\mathcal{F}_T}
\newcommand\GTX{\mathcal{G}_{TX}}
\newcommand\GTXX{\mathcal{G}_{TXX}}
\newcommand\FTX{\mathcal{F}_{TX}}

\newcommand\OmB{\Lambda}   
\newcommand\YB{\Pi}        
\newcommand\DB{\Sigma}     
\newcommand\EB{\Upsilon}   

\numberwithin{equation}{section}
\numberwithin{figure}{section}
\numberwithin{table}{section}

\newcolumntype{Y}{>{\centering\arraybackslash}X}
\pgfplotsset{compat=1.6}

\author{Simon Iteanu$^{a}$ \\
\small $^{a}${\it{Universit\'e Paris-Saclay, CNRS/IN2P3, IJCLab, 91405 Orsay, France}}\\
}

\date{}
\title{Dipolar  Perturbations of  Black Holes in DHOST Theories}

\begin{document}

\maketitle

\begin{center}
{\bf Abstract}
\end{center}

We study the dipolar sector of polar perturbations of static black holes with primary hair. These solutions arise in a subfamily of degenerate higher-order scalar-tensor (DHOST) theories, and this work extends the analysis initiated for the monopole in \cite{Iteanu2026rph}.
We recast the equation of motion of the dipolar degree of freedom as a Schr\"odinger-like wave equation, and show that the associated effective metric of the dipole, defined from spatial infinity down to the central singularity, is identical to that of the monopole. We also confirm that this effective metric is preserved by disformal transformations.
For a class of hairy black holes, we show that, depending on the boundary condition imposed at the central singularity, the spectrum of both the monopole and the dipole is either positive or contains exactly one negative eigenvalue. The transition occurs at a critical boundary condition, characterised by the existence of a static zero mode, which we determine explicitly for the monopole and numerically for the dipole.
We finally examine stealth black hole solutions, with either a non-constant or a constant kinetic density, as well as the black hole of 4D Einstein-Gauss-Bonnet (4DEGB) gravity. For the stealth solution with a non-constant kinetic density and for the 4DEGB black hole, the effective metric degenerates and there is no propagating dipolar degree of freedom at linear order. For the stealth solution with a constant kinetic density, we identify a stable dipolar degree of freedom, except for special values of the coupling constants of the theory.

\newpage
\tableofcontents
\newpage

\section{Introduction}
This paper is a companion to \cite{Iteanu2026rph}, where we investigated the radial perturbations of static black holes in a subclass of Degenerate Higher-Order Scalar-Tensor (DHOST) theories. We developed a systematic procedure to isolate the propagating degree of freedom and recast its equation of motion into a Schrödinger-like form. Applying this formalism to the hairy black hole solutions of \cite{Bakopoulos:2023fmv,Baake:2023zsq,Bakopoulos:2023sdm,Charmousis:2025xug}, we found that the time coordinate adapted to the Schrödinger equation coincides, up to a normalization, with the background scalar field. Remarkably, the associated radial coordinate extends across the event horizon into the black hole interior, down to the central singularity, while spanning a finite interval, in contrast with the usual tortoise coordinate of General Relativity (GR). This structure has important consequences for the boundary conditions and the stability analysis. In particular, we established the existence of a positive self-adjoint extension of the Schrödinger operator, ensuring the linear stability of the monopole, and investigated the corresponding spectrum of normal modes.
Since the endpoint at the singularity is of limit-circle type, this extension is however not unique, and the stability of the monopole under other admissible boundary conditions was left open.

We also applied our formalism to two classes of stealth black holes. For the solution with a nonconstant kinetic density $X$, studied in \cite{Sirera:2024ghv}, we found that no propagating monopole degree of freedom arises at linear order. By contrast, for the stealth solution with constant $X$ considered in \cite{Takahashi:2021bml}, we established the linear stability of the monopole in the appropriate parameter regime.

In the present paper, we extend our analysis of polar perturbations to the dipolar sector, $\ell=1$. We show that the Schrödinger coordinates governing the dipolar mode coincide with those obtained for the monopole, as suggested by the results of \cite{Takahashi:2021bml}, so that the monopole and dipole modes propagate along the same effective light cone.
\footnote{From the principal symbol of the master equation, one can immediately read off the coefficients of the inverse effective metric $\tilde g^{\mu\nu}$, up to a global factor. In 1+1 dimensions, this conformal class of metrics is entirely characterised by its light cone, and we will refer to it as the effective metric.}
For the hairy black holes, we establish the stability of the dipole for the Friedrichs extension, i.e., the self-adjoint extension of the Schr\"odinger operator which selects the most regular solutions at the endpoints. 
We then show that, for any admissible boundary condition at the central singularity, both the monopole and the dipole have at most one unstable mode. 

The dipolar analysis also represents an important step toward understanding the full system of polar perturbations for higher multipoles, $\ell\geq 2$, where the scalar and tensor degrees of freedom are generically coupled. In particular, we expect the two master variables describing the higher multipolar sector to be related to linear combinations of those identified in the monopole and dipole sectors as shown in \cite{Takahashi:2021bml,Kobayashi:2014wsa}.

The paper is organized as follows. In the next section, we introduce our general framework and present our systematic procedure for extracting the degree of freedom associated with dipolar oscillations and formulating its equation of motion in a Schr\"odinger-like form, giving explicit expressions for a general quadratic Horndeski theory in Appendix~\ref{app:S_Horndeski}. Section~\ref{Sec:HairyBH} applies our procedure to study the dipolar oscillations of the solutions found in \cite{Bakopoulos:2023fmv}, and analyses, for both the monopole and the dipole, the stability under generic boundary conditions at the central singularity. Section~\ref{section:Non_propagating} is devoted to two Horndeski backgrounds on which the dipole does not propagate: the stealth black hole with a non-constant $X$ and the black hole of 4D Einstein-Gauss-Bonnet (4DEGB) gravity. Section~\ref{stealth section} focuses on stealth black holes with a constant $X$, for which we establish the stability of the dipole.
The final section summarizes our results. The appendices contain the gauge transformations of the perturbations (App.~\ref{app:Component_transfo}).
All of the computations in this paper can be found on a GitHub page where the full expressions for all these quantities can be downloaded in their most general form \cite{Iteanu2026Git}.

\section{Theoretical framework and dipolar perturbations}
In this section, we present the general procedure to compute, from the perturbation equations about an arbitrary background in DHOST theories, a Schr\"odinger-like equation for the dipolar perturbations.

\subsection{DHOST theories and background solutions}
We consider  DHOST theories (up to quadratic terms in the second derivatives of the scalar field) that are described by the action \cite{Langlois:2015cwa,Langlois:2015skt}
\begin{align}
\label{DHOSTaction}
 S\left[g_{\mu \nu}, \phi\right]= \int \mathrm{d}^4 x \sqrt{-g} \left( P(X, \phi)+ Q(X, \phi) \square \phi+\Ftwo(X, \phi) R +  \sum_{i=1}^5 A_i(X, \phi) L_i^{(2)} \right)  , 
\end{align}
where the kinetic density is defined by
\begin{equation}
    X\equiv  -\frac{1}{2} \partial_\mu \phi \, \partial^\mu \phi\,,
\end{equation}
and  where the five elementary quadratic Lagrangians are defined by
\cite{Langlois:2015cwa} 
\begin{eqnarray*}
&L_{1}^{(2)}=
\phi^{\mu\nu}\phi_{\mu\nu}\,, \qquad L_{2}^{(2)}=\left( \square\phi\right)^{2}\,, \qquad L_{3}^{(2)}= (\square\phi)\phi^{\mu}\phi_{\mu\nu}\phi^\nu\,,
\qquad \\
  & L_{4}^{(2)}=\phi^\lambda\phi_{\lambda\mu}\phi^{\mu\nu}\phi_\nu \,, \qquad \qquad
  L_{5}^{(2)}=\left(\phi^{\mu}\phi_{\mu\nu}\phi^\nu\right)^{2}\,,
  \qquad
\end{eqnarray*}
using the  notation $\phi_\mu\equiv \nabla_{\mu}\phi$ and $\phi_{\mu\nu}\equiv \nabla_{\nu}\nabla_{\mu}\phi$.

The theories that will be discussed in this paper all belong to a sub-class of
DHOST theories, known as Beyond Horndeski theories~\cite{Gleyzes:2014dya,Gleyzes:2014qga}, characterised by the conditions
\begin{eqnarray}
\label{eq:Ai_GLPV}
A_2=-A_1   \,, \qquad A_4=-A_3= \frac{A_1+\Ftwo_{X}}{X} \,, \qquad A_5=0\,.
\end{eqnarray}
This sub-class contains the Horndeski theories~\cite{Horndeski:1974wa}, which satisfy the more restrictive conditions
\begin{eqnarray}
\label{Ai_Horndeski}
A_2=-A_1= \Ftwo_{X}  \,, \qquad A_3=A_4=A_5=0\,.
\end{eqnarray}
From now on, we restrict our attention to Lagrangians in which the functions $P$, $Q$, $\Ftwo$ and $A_1$ depend only on $X$, which implies in particular that the theory is invariant under the shift symmetry $\phi \rightarrow \phi + c$ (where $c$ is a constant).
In the rest of the paper, we further set $Q=0$, so that the theory is also invariant under the parity symmetry $\phi \rightarrow -\phi$. 

The backgrounds which we shall be perturbing are static and spherically symmetric solutions to the equations of motion,
\begin{align}
\label{eq:line_element}
\mathrm{d}\bar{s}^{\,2} &=-\cA(r)\mathrm{d}t^2+\frac{\mathrm{d}r^2}{\cB(r)}+r^2\mathrm{d}\Omega^2,\qquad \mathrm{d}\Omega^2=\mathrm{d}\theta^2+\sin^2\theta\,\mathrm{d}\varphi^2\,,
\end{align}
with a  scalar field admitting a linear time dependence (allowed by the shift symmetry),
\begin{equation}
\label{phiback}
    \phiback=q t+\psi(r) \, .
\end{equation}

\subsection{Perturbation theory and gauge fixing}
We now consider linear perturbations about the background solution \eqref{eq:line_element}-\eqref{phiback}. Let $h_{\mu \nu}$ be the perturbation about the background metric $\gmunuback$ and $\delta \phi$ the scalar perturbation about the background scalar field  $\phiback$:
\begin{equation}
   h_{\mu \nu} =g_{\mu \nu}-\gmunuback, \qquad \delta \phi=\phi-\phiback.
   \label{pertgeneral}
\end{equation}
 Due to the spherical symmetry of the background, we project both perturbations on the basis of spherical harmonics $Y_{\ell m}(\theta, \varphi)$, and separate them into axial (odd-parity) and polar (even-parity) modes, as the two sectors are decoupled at linear order.
We have already  studied axial modes as well as the radial mode in \cite{Noui:2023ksf,Iteanu2026rph}.
For a general $\ell\geq2$, the polar modes 
are described  by seven components of the metric perturbation and the perturbation of the scalar field. In this article we study the dipole $\ell=1$,  where the  non-trivial components of the perturbations \eqref{pertgeneral} can be parametrized as follows,
\begin{align}
&h_{t t}  =\cA(r)  H_0(t, r)Y_{\ell m}(\theta, \varphi) \, ,\quad
h_{t r} = H_1(t, r)Y_{\ell m}(\theta, \varphi) \, , \quad 
h_{r r}  =\frac{1}{\cA(r)}  H_2(t, r)Y_{\ell m}(\theta, \varphi) \, , \\[6pt]
& h_{a b}= K(t, r) \overline{g}_{a b} Y_{\ell m}(\theta, \varphi)+ G(t, r) \nabla_a \nabla_b Y_{\ell m}(\theta, \varphi)\quad (a,b \in \{\theta,\varphi \}) \, ,\\[6pt]
& h_{t a}= \beta(t, r) \partial_a Y_{\ell m}(\theta, \varphi), \quad h_{r a}= \alpha(t, r) \partial_a Y_{\ell m}(\theta, \varphi)\, ,\\[6pt]
&
\delta \phi= \delta \phi(t, r) Y_{\ell m} \, .
\end{align}

As described in Appendix~\ref{app:Component_transfo}, for $\ell=1$, only the linear combination $r^2K-G$ enters $h_{ab}$ and should be treated as a single perturbation variable. Using the transformation rules listed in this appendix, we choose the gauge
\begin{equation}
\label{gauge}
\delta \phi=0 \, ,\qquad 
r^2K-G=0 \, , \qquad
H_0=\frac{2 \psi' }{q\, r}\beta\, ,
\end{equation}
where we assume $q\neq0$. 
This gauge is complete\footnote{The first two gauge choices fix $T=-R\,\psi'/q$ and $\Theta=r R$.
Then the variation of $H_0-\frac{2\psi'}{qr}\beta$ under a gauge transformation reduces to $R\left(\frac{\cA'}{\cA}+\frac{2\cA\psi'^2}{q^2 r}\right)$,
where the terms involving derivatives of $R$ vanish. Hence no residual gauge freedom is left as long as this coefficient does not vanish, which is the case for all the backgrounds considered here.}\cite{Motohashi_2016}.
In what follows, a prime denotes the derivative with respect to $r$ and a dot the derivative with respect to $t$.
Finally, after the gauge fixing, we are left with only four independent fields $\alpha,\beta,H_1$ and $H_2$.

\subsection{Master equation for the dipole perturbation}
\label{perturbations section}
Let us now outline the strategy for the construction of the master variable describing the dipole perturbation and of the corresponding second-order equation.
We give here the scheme for a generic quadratic Horndeski theory (with $X'\neq0$) described by the system of background equations given in eqs.~(2.7)-(2.9) of \cite{Charmousis:2025xug}. The same resolution can be applied, for instance, to Beyond Horndeski theories, up to slight variations in the definitions of the various variables involved.
In the rest of this section, we use the notations introduced in Sec.~4.1 of \cite{Iteanu2026rph}.

Among the ten linearized metric equations $\mathscr{E}_{\mu\nu}=0$, many are trivially satisfied, and we consider the five equations $\mathscr{E}_{tt}$, $\mathscr{E}_{tr}$, $\mathscr{E}_{rr}$, $\mathscr{E}_{t\theta}$ and $\mathscr{E}_{r\theta}$\footnote{The scalar field equation, as well as the remaining non-vanishing metric equations, are related to these five equations and their derivatives through the Bianchi identities. We have checked explicitly in \cite{Iteanu2026Git} that all the equations are satisfied by the solution constructed below.}.
These five equations are not independent: as we will see below, $\mathscr{E}_{r\theta}$ is redundant with the four others. We nevertheless keep it, as it greatly simplifies the resolution of the system.
These equations are linear combinations of the fields mentioned above and their derivatives, and take the schematic form
\begin{eqnarray}
\mathscr{E}_{tt} & = & [\alpha,\beta ,H_1 ,H_2,\alpha',H_1',H_2',\dot{H_2}] \, ,\label{tt_equation} \\
\mathscr{E}_{tr} & = & [\alpha,\beta ,H_1,H_2,\beta',\dot{\alpha},\dot{H_2}] \, , \label{tr_equation} \\
\mathscr{E}_{rr} & = & [\alpha,\beta ,H_1,H_2,\beta',\dot{\beta},\dot{H_1}] \, , \label{rr_equation}\\
\mathscr{E}_{t\theta} & = & [\alpha,\beta ,H_1,H_2,H_1',\dot{\alpha},\dot{H_2},\beta^{\prime \prime},\dot{\alpha}']  \, ,\label{ttheta_equation}\\
\mathscr{E}_{r\theta} & = & [\alpha,\beta ,H_1,H_2,\beta',\dot{\beta},\dot{H_1},\dot{\beta}',\ddot{\alpha}] \, ,\label{rtheta_equation}
\end{eqnarray}
where, here and throughout, the notation $[\cdots]$ denotes a linear combination of the terms enclosed within the brackets, with coefficients that depend on the background solution. Their general explicit expressions are not needed here.

Second-order derivatives only appear in the angular equations $\mathscr{E}_{t\theta}$ and $\mathscr{E}_{r\theta}$, through the radial and time derivatives of the combination $\dot{\alpha}-\beta'$ respectively. We therefore introduce the auxiliary field
\begin{equation}
\vv= \dot{\alpha}-\beta'+[\alpha, \beta, H_1]\,,
\label{nnvv}
\end{equation}
where the coefficients of the linear combination $[\alpha,\beta,H_1]$ are uniquely fixed by requiring that, once $\dot\alpha$ is expressed using \eqref{nnvv}, $\dot\beta$, $\beta'$ and $\dot H_1$ drop out of $\mathscr{E}_{r\theta}$. Both angular equations then become of first order and take the form
\begin{eqnarray}
\mathscr{E}_{t\theta} & = & [\vv,H_2,\alpha ,\beta ,H_1,\alpha',H_1',\vv',\dot{H_2}] \, , \\
\mathscr{E}_{r\theta} & = & [\vv,H_2,\alpha ,\beta ,H_1,\dot{\vv}] \, .
\end{eqnarray}
Let us stress that the definition \eqref{nnvv} must be understood as an additional differential constraint ensuring the equivalence with the original system.

\noindent Next, we replace the fields $H_1$ and $H_2$ by the new fields $\zv$ and $\nv$,
\begin{eqnarray}
\zv&= &[H_1,\vv,\alpha,\beta]\, , \\
\nv&=&[H_2,H_1,\alpha,\beta]\, ,
\end{eqnarray}
and combine the equations \eqref{tt_equation}-\eqref{rtheta_equation} in such a way that the whole system takes the much simpler form
\begin{eqnarray}
\mathscr{E}_{1} & = & [\alpha ,\vv,\zv,\nv,\nv'],\\
\mathscr{E}_{2} & = & [\beta ,\vv,\nv,\zv,\dot{\nv}],\\
\mathscr{E}_{3} & = & [\alpha ,\beta ,\nv,\zv,\vv,\dot{\vv},\dot{\zv}] \,,\\
\mathscr{E}_{4} & = & [\alpha ,\nv,\zv,\vv,\zv'] \, ,\\
\mathscr{E}_{5} & = & [\alpha ,\beta ,\nv,\zv,\vv,\dot{\vv}] \, ,
\end{eqnarray}
where $\mathscr{E}_3$ and $\mathscr{E}_5$ correspond to $\mathscr{E}_{rr}$ and $\mathscr{E}_{r\theta}$ respectively.
The system $\{\Ne=0 \, ,\alg=0,\mathscr{E}_{4}=0\}$ can be inverted to express $\alpha$, $\beta$ and $\vv$ in terms of $\zv$, $\nv$ and their derivatives,
\begin{eqnarray}
\label{eq:vv_symbolicsolve}
\vv&=&[\zv,\nv,\nv',\zv']\, ,\\
\alpha &=&[\zv,\nv,\nv',\zv']\,,
\label{eq:alpha_symbolicsolve}\\
\beta &=&[\zv,\nv,\nv',\zv',\dot{\nv}] \, .
\end{eqnarray}
Once these expressions are substituted, the two remaining equations $\mathscr{E}_3$ and $\mathscr{E}_5$ both involve $\zv$, $\nv$, their first derivatives and the mixed derivatives $\dot{\zv}'$ and $\dot{\nv}'$. However, there exists a linear combination of them in which all the radial and mixed derivatives cancel,
\begin{equation}
\mathscr{E}_3 + a(r)\, \mathscr{E}_5 = [\zv,\nv,\dot{\nv},\dot{\zv}] \, .
\label{eq:E3_ZN}
\end{equation}
We finally introduce the master variable $\mv$, defined as a linear combination of $\nv$ and $\zv$ with coefficients chosen such that all the time derivatives in \eqref{eq:E3_ZN} combine into $\dot{\mv}$:
\begin{equation}
\label{eq:def_sigma}
\mv= [\nv,\zv]\,, \qquad \mathscr{E}_3 + a\, \mathscr{E}_5=[\mv,\zv,\dot{\mv}]\,.
\end{equation}
Solving this equation for $\zv$, one obtains
\begin{equation}
\label{eq:zv_symbolicsolvesigma}
\zv= [\mv,\dot\mv]\, ,
\end{equation}
and one can check that $\mathscr{E}_3$ and $\mathscr{E}_5$ are then separately satisfied. Hence, once the other equations are solved, $\mathscr{E}_{rr}$ and $\mathscr{E}_{r\theta}$ reduce to a single equation, which illustrates the redundancy mentioned above.

Finally, using Eqs.~\eqref{eq:def_sigma} and \eqref{eq:zv_symbolicsolvesigma}, we can express all the original metric perturbations in terms of the master variable $\mv$ and its derivatives. Substituting these expressions into the defining relation \eqref{nnvv} then yields a second-order differential equation for $\mv(t,r)$ of the form
\begin{equation}
\label{def_second_order_equation}
d_1 \, \mv''+d_2\, \ddot{\mv}+d_3 \, \dot{\mv}'+d_4 \, \mv'+d_5 \, \dot{\mv}+d_6 \, \mv=0 \, .
\end{equation}
The coefficients $d_i$ can be computed from the coefficients entering the metric equations \eqref{tt_equation}-\eqref{rtheta_equation}, but their expressions are too cumbersome to be written here in full generality. We give their explicit expressions in the following sections for some cases of interest.
\subsection{Schr\"odinger-like equation}
\label{sec:Schro_equation}
In order to rewrite the master equation \eqref{def_second_order_equation} in the form of  a Schr\"odinger-like equation, we introduce new time and radial variables:
\begin{align}
\label{def_tstar}
    t_*&=t +\int W(r)\mathrm{d}r \,, \qquad W=-\frac{d_3}{2d_1} \,, \\[7pt]
    r_*&=\int \frac{\mathrm{d}r}{n(r)}\,, \qquad n =\sqrt{ - \frac{d_1}{d_{2*}}}\,, \, \quad \text{with} \quad d_{2*}\equiv d_2+ W^2 d_1 + W d_3  \, ,
    \label{def_rstar}
    \end{align}
which requires $d_1/d_{2*}<0$, or equivalently $d_3^2-4d_1d_2>0$ (assuming $d_1\neq0$), corresponding to the hyperbolicity condition of the second-order equation.
From now on, we will use the notation $\partial_*={\partial}/{\partial r_*}$.
Let us mention here that, even though the coordinates of the Schr\"odinger equation for the dipole take a similar form to those of the monopole, there is no reason a priori for them to be equal.
Nonetheless, for all the Horndeski theories, both $t_*$ and $n$ for the dipole are equal to their counterparts for the monopole, once the background equations are imposed,
\begin{align}
        W_{\ell=1}=W_{\ell=0}=W\, , \qquad
    n_{\ell=1}=n_{\ell=0}=n\,,
\label{eq:SametandNHorn}
\end{align}
confirming that generically the monopole and dipole perturbations share the same effective light cone.
The explicit expressions of \eqref{eq:SametandNHorn} for all Horndeski theories with $X'\neq0$ are given in Appendix~\ref{app:Schro_horndeski}. 
Note also that this result extends to any theory related to a quadratic Horndeski theory by an invertible disformal transformation \cite{Zumalac_rregui_2014,BenAchour:2016cay}, which includes the Beyond Horndeski theories of Sec.~\ref{Sec:HairyBH}.

In these coordinates, the coefficient of $\partial_{t_*}\mv$ also vanishes once the background equations are imposed.
We further introduce a normalization of the function $\mv$, $\smv=\mv/{\cal N}$, where ${\cal N}$ satisfies
\begin{equation}
\label{normalisation}
d_1 \left(\frac{{\cal N}^2}{n}\right)' + d_4 \,\frac{{\cal N}^2}{n}=0 \, ,
\end{equation}
so that no first-order derivative with respect to $r_*$ remains, and we obtain the wave equation
\begin{eqnarray}
\label{waveop}
    \left[\frac{\partial^2 }{\partial t_*^2} - \frac{\partial^2 }{\partial r_*^2}\right] \smv + \Veff \, \smv \; = \; 0 \,,
\end{eqnarray}
where the master variable $\smv$ plays the role of the variable $\Psi$ introduced in Ref.~\cite{Iteanu2026rph} for the monopole. The effective potential is given by
\begin{eqnarray}
\label{Veffgeneral}
    \Veff = \frac{1}{d_{2*}} \left( \frac{{\cal N}''}{\cal N} d_1 + \frac{{\cal N}'}{\cal N} d_4 + d_6 \right) \, .
\end{eqnarray}

Finally, going from the time domain to the frequency domain via  a Fourier transform  with respect to the new time coordinate $t_*$, i.e. 
\begin{equation}
   \smv(t_*,r) = \smv(r) e^{-i\omega t_*} \, ,
    \label{Ftransform}
\end{equation}
the  equation for $\smv$ takes the familiar Schr\"odinger-like form
\begin{equation}
\label{Schrödinger_equation_generic}
\Schroop\smv=-\frac{\mathrm{d}^2 \smv}{\mathrm{d}r_*^2}+ \Veff \, \smv \; = \; \omega^2 \, \smv \, ,
\end{equation}
where $\Schroop$  denotes the Schr\"odinger operator and $\Veff$ should be viewed as a function of $r_*$, even if it is given explicitly only in terms of $r$. 

In all the cases considered in this paper where the dipole propagates, we will show that the potential can be written as
\begin{equation}
\label{eq:S_decomp_Vs}
\Veff=S^2-\partial_*S+\Vs\,,
\end{equation}
where $S$ is a function regular on the whole interval where $r_*$ is defined, and $\Vs\geq0$ in the region of parameter space considered.
As shown in Sec.~\ref{Sec:HairyBH}, such a decomposition implies the positivity of the Schr\"odinger operator, up to boundary terms. 
In Appendix~\ref{app:Schro_horndeski}, we also give a compact formula for $S$ for any solution of quadratic Horndeski theory, for both the monopole for which $\Vs=0$ and for the dipole with a specific choice of $\Vs$.

\section{Hairy black holes in Beyond Horndeski theories}
\label{Sec:HairyBH}
\subsection{Black hole solutions}
To start our stability analysis, we first focus on the black hole solutions obtained for a family of  DHOST theories \cite{Bakopoulos:2023fmv} (see also \cite{Baake:2023zsq,Bakopoulos:2023sdm,Charmousis:2025xug,bahamonde2026nonreciprocalblackholesprimary}), within the Beyond Horndeski subclass. These theories are characterised by the  functions 
\begin{align}
\label{P_n}
    P(X)&=-\frac{2 \alpha}{ \lambda^2} X^\nth \, , \quad
    \Ftwo(X)=1- 2X A_1(X)\, , \quad
    A_1(X) =\frac{\alpha}{2} X^{\nth-1}\,, \quad 
A_3(X)=\frac{\alpha}{2}(2\nth-1) X^{\nth-2}\,,
\qquad
\end{align}
while $A_2=-A_1$, $A_4=-A_3$ and $A_5=0$. They depend on three parameters: $\nth$, which, for simplicity, is assumed to take positive integer or half-integer values henceforth; $\alpha$, a dimensionless coupling constant; and $\lambda$, a coupling constant with the dimension of length. 
We treat the same cases considered in \cite{Iteanu2026rph} and consider $\nth>3/2$.

These theories admit static  spherically symmetric solutions  of the form \eqref{eq:line_element}-\eqref{phiback},
with the  scalar field profile given by
\begin{equation}
X=\frac{q^2 /2}{1+(r / \lambda)^2}\,, \qquad \psi^{\prime}(r)^2=\frac{q^2}{\cA(r)^2}\left[1-\frac{\cA(r)}{1+(r / \lambda)^2}\right] \,, \label{scalar}
\end{equation}
while the metric components read
\begin{equation}
    \cA=\cB=1-\frac{2\mu}{r}-\xi_\nth\frac{2\lambda}{r} \,\Xi_\nth(r/\lambda)\,.
    \label{solAq0}
\end{equation}
In the above expression, $\mu$ is an integration constant (with the dimension of a length) and we have introduced the notations
\begin{equation}
\label{xi}
    \xi_\nth\equiv \alpha ({2\nth-1}) \left({q^2}/{2}\right)^{\nth} \,, \qquad  \Xi_\nth(x)\equiv\int_0^x \mathrm{d}u \; \frac{u^2}{\left(1+u^2\right)^\nth} \,.
\end{equation}

For $\nth>3/2$, $\Xi_p(r/\lambda)$ tends to a constant at spatial infinity, which therefore  contributes to  the ADM mass $M$\footnote{Note that $M$ has the dimension of a length, since we work implicitly in units where $G=1$ and $c=1$.} 
\begin{equation}
\label{mass}
    M\equiv \mu+\frac{\sqrt{\pi}\,\Gamma(\nth-\frac32)}{4\, \Gamma(\nth)}\xi_\nth\, \lambda\equiv \mu+\mreg\qquad (\nth>3/2)\,.
\end{equation}
For $\nth\leq3/2$, the integral defining $\Xi_\nth(\infty)$ diverges and the metric is not asymptotically flat (for instance $\cA\to1-2\xi_1$ when $\nth=1$).
Let us set henceforth $\lambda$ to unity in order to simplify expressions. We will restore it if necessary later on.
The resulting metric describes a black hole geometry if $\cA$ vanishes for some finite radius $\rh$. Otherwise the solution describes a naked singularity (or a soliton if the metric is regular at the origin, i.e. when $M=\mreg$, or, equivalently, $\mu=0$).  
We have plotted $\cA$ in the left column of Fig.~\ref{fig:A_V_different_p} for different values of $\nth$ and $M/\mreg$. For further details and plots of $\cA$, the interested reader should refer to \cite{Noui:2023ksf,Iteanu2026rph}.

\subsection{Schr\"odinger-like reformulation}
Following the resolution procedure described above, we obtain a second-order equation that we turn into a  Schr\"odinger equation
by introducing the new time coordinate defined in \eqref{def_tstar}: 
\begin{equation}
\label{tstar_p=2}
t_*=t+\frac1q\psi=\frac\phiback q\,.
\end{equation}
Similarly to the monopole, $t_*$ coincides with the scalar field, up to a rescaling by $1/q$. Then, for the radial coordinate,
\begin{align}
\label{r_star_hairy}
   r_*(r)\equiv \int^r \frac{\mathrm{d}x}{n(x)}\,, \qquad  n(r)&=\frac{\left(r^2+1\right) \sqrt{r^2+1-\cA(r)}}{\sqrt{2 p-1}\ r}\,.
 \end{align}
Here again we obtain the same function $n$ as for the monopole, which shows that the monopole and the dipole share the same radial sound speed, even beyond Horndeski theories.

An in-depth analysis of this function $n$ is given in \cite{Iteanu2026rph}. Let us just mention here that if $\mu=M-\mreg>0$ \eqref{mass}, then $n(r)$ is defined over the whole interval $]0,+\infty[$, from the central singularity at $r=0$ all the way to spatial infinity. In this case, the lower bound in the above integral defining $r_*$ can be chosen to be zero, and the integral converges both for $r\to 0$ and $r\to +\infty$, meaning that $r_*$ is defined on a compact interval of the form $[0,\rmax]$ 
\footnote{Let us mention that the convergence at infinity follows from the behavior of $n$ at spatial infinity:
$ n=\frac{1+r^2}{\sqrt{2\nth-1}}\left[1+{\cal O}(r^{-3})\right].$
It grows as $r^2$, like the function $1+r^2$ of anti-de Sitter spacetime, so that the boundary $r\to+\infty$ is reached at a finite value $\rmax$ of the Schr\"odinger coordinate.}.
Unless stated otherwise, we focus on this case in the following.

\medskip

Let us mention that the above result extends to any theory related by a regular and invertible disformal transformation, as expected from the invariance of the vacuum physics under such field redefinitions. As an explicit example, we derived the Schr\"odinger equation governing the dipolar perturbations of Horndeski theories obtained through a disformal transformation (see Refs.~\cite{Iteanu2026rph,Charmousis:2025xug}), using the equations in Ref.~\cite{Iteanu2026Git}. The same equivalence had previously been established for the monopolar sector in Ref.~\cite{Iteanu2026rph}.
 
\subsubsection{Effective potential}
In the Schr\"odinger coordinates $(t_*,r_*)$, the  effective potential which appears in the Schr\"odinger-like equation \eqref{Schrödinger_equation_generic} can be written as
\begin{eqnarray}
    \Veff \; = \; S^2 - \partial_* S + \Vs \,  ,
    \label{Veff_HairyBH}
\end{eqnarray}
where we have defined 
\begin{equation}
  S=-\partial_*\ln\!\left[\frac{\sqrt{n(r)}\;r\,(1+r^2)^{\frac{p-1}{2}}}{\big|3(\cA-1)(1+r^2)^p+2\xi_{\nth} r^2\big|}\right],
  \qquad
  \Vs=\frac{1-\cA}{2r^2} \,.
  \label{SandVsdef_Hairy}
\end{equation}
Note already that $S$ is regular on $r>0$ and that $\Vs>0$ whenever $M>\max \left(0, \mreg\right)$.
Recall also that for the monopole $\ell=0$, we had found
\begin{equation}
\label{Sfunction_Monopole_Hairy}
\smono(r)=-\partial_*\ln\!\left[
\frac{(1+r^2)^{p/2}
\left|\Xi_p(r)+c_S\right|}
{r\sqrt{n(r)}}
\right] \, , \qquad \Vs=0 \, ,
\end{equation}
where $c_S$ is an integration constant that we had set to $0$ in Ref.~\cite{Iteanu2026rph}. This choice for $c_S$ amounted to picking one of the two branches for the zero mode of the equation. 

\begin{figure}[!t]
    \centering
    \begin{subfigure}[t]{0.49\textwidth}
        \centering
        \includegraphics[width=\textwidth]{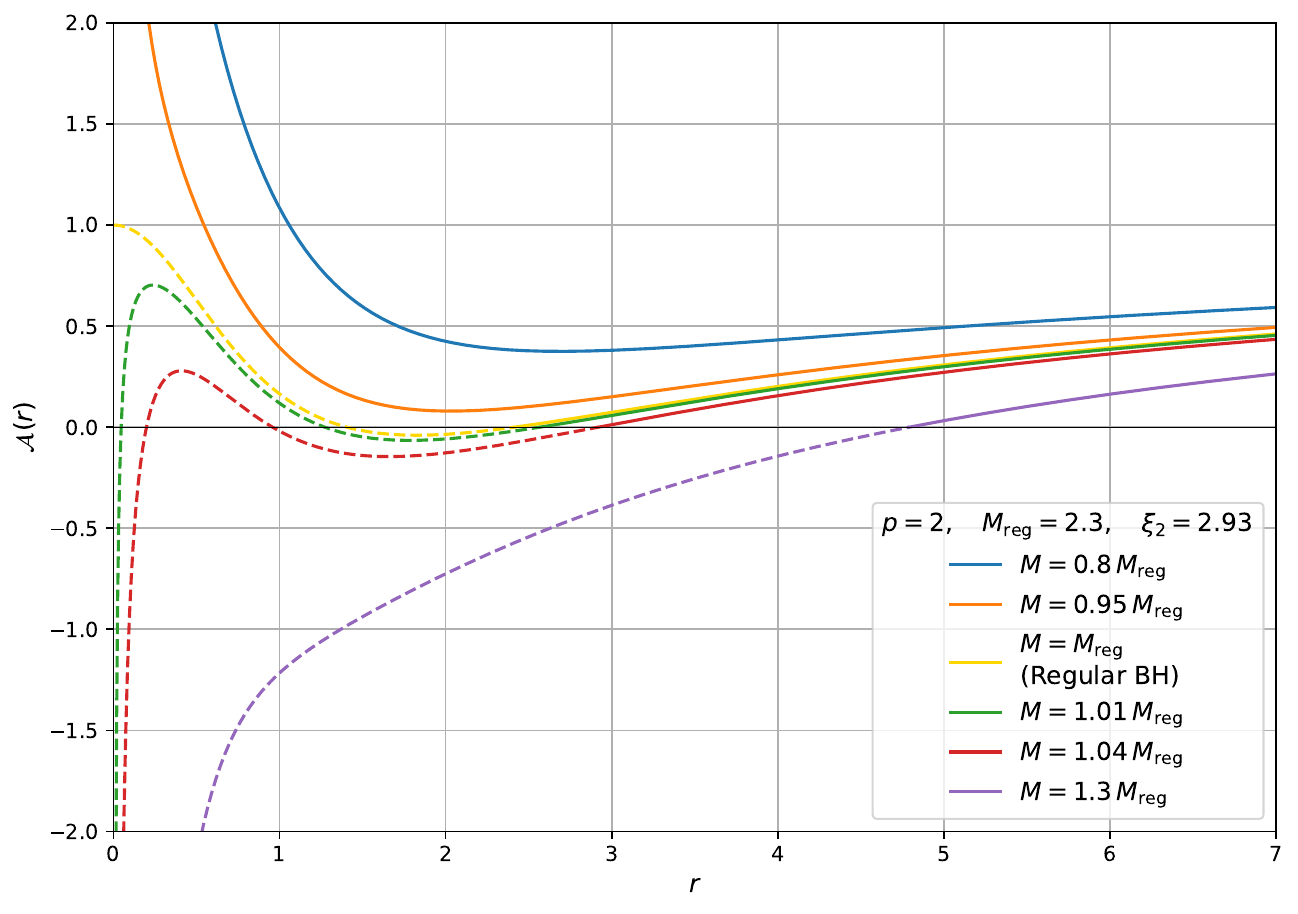}
    \end{subfigure}
    \begin{subfigure}[t]{0.49\textwidth}
        \centering
        \includegraphics[width=\textwidth]{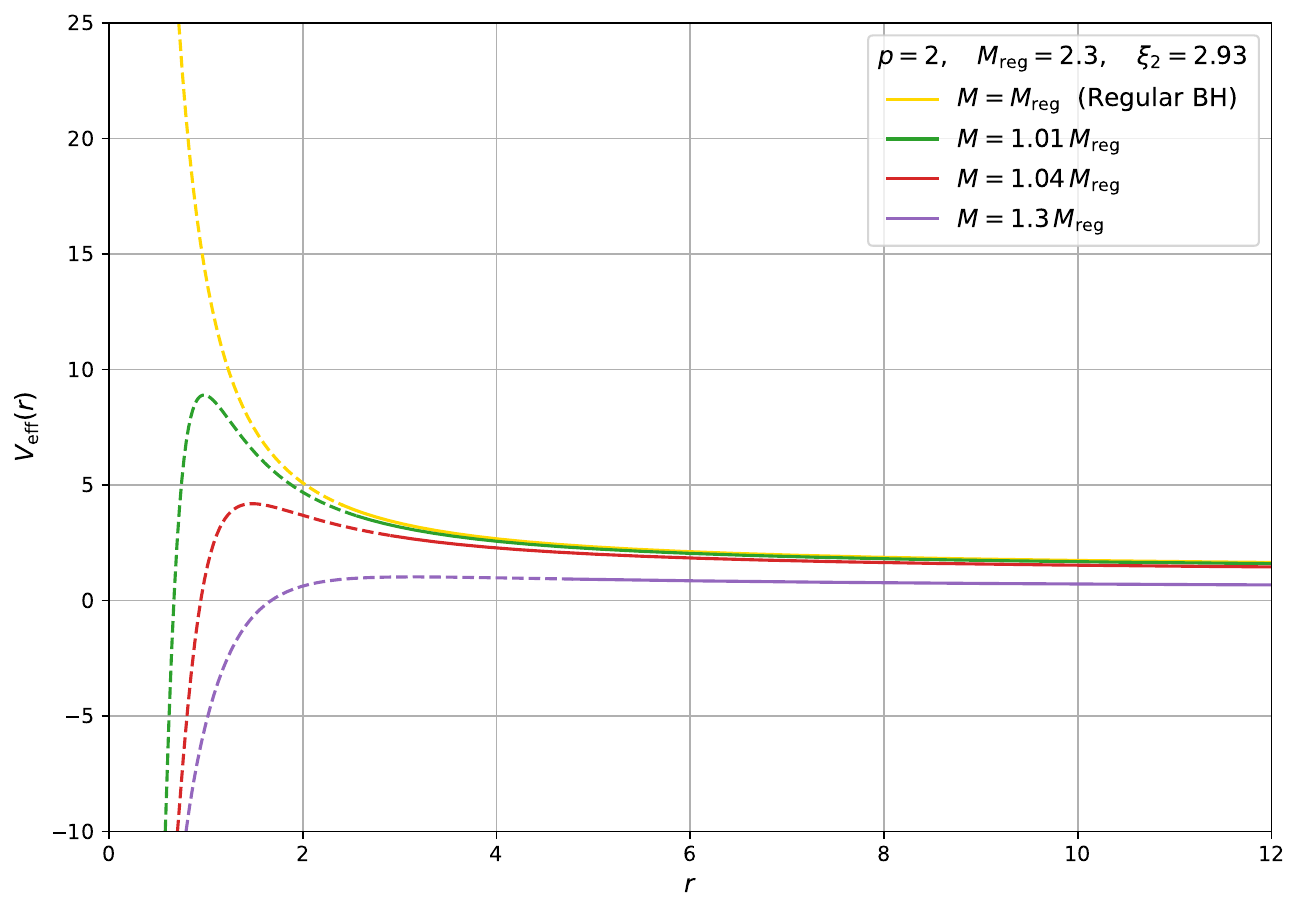}
    \end{subfigure}

    \begin{subfigure}[t]{0.49\textwidth}
        \centering
        \includegraphics[width=\textwidth]{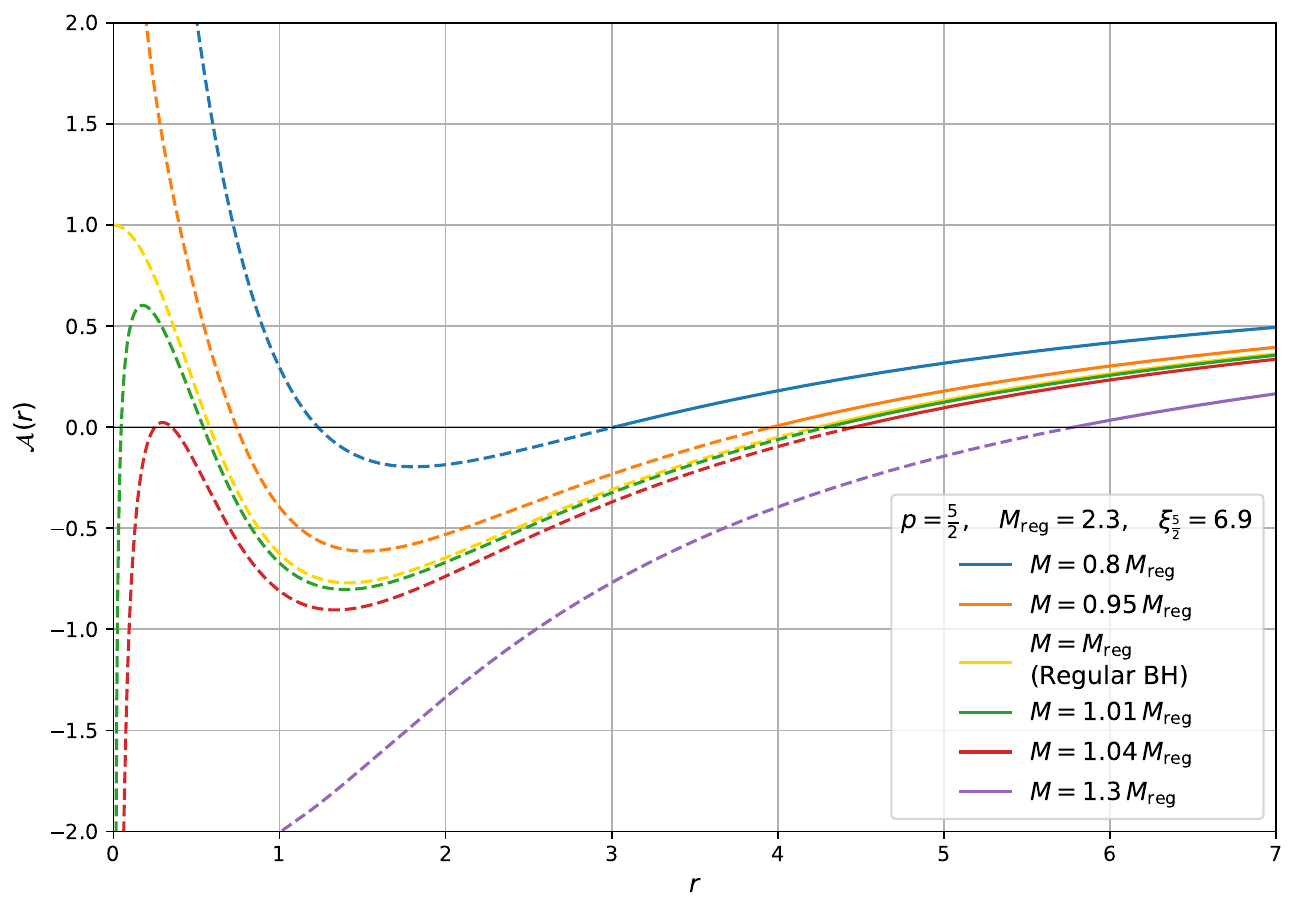}
    \end{subfigure}
    \begin{subfigure}[t]{0.49\textwidth}
        \centering
        \includegraphics[width=\textwidth]{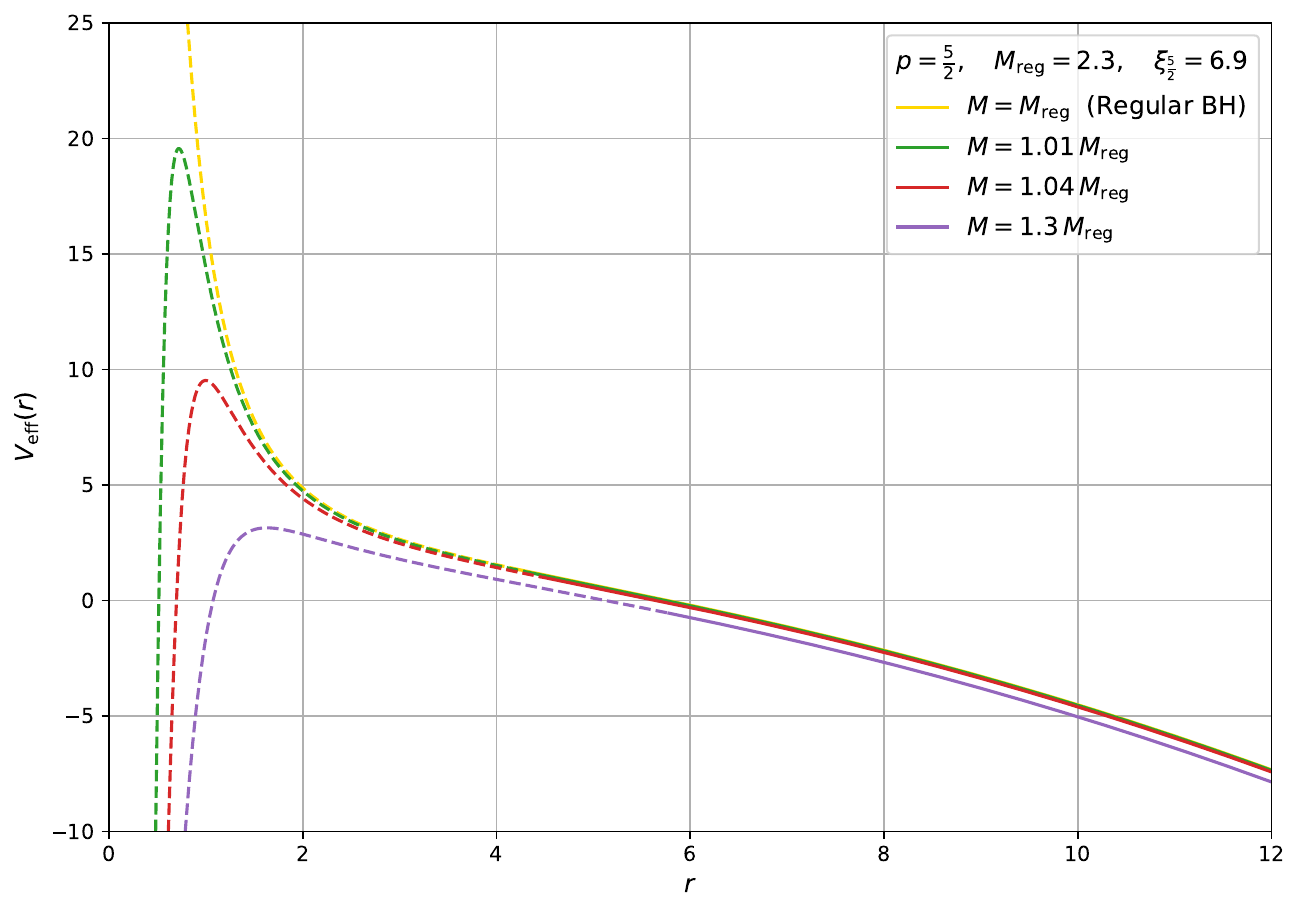}
    \end{subfigure}
    \begin{subfigure}[t]{0.49\textwidth}
        \centering
        \includegraphics[width=\textwidth]{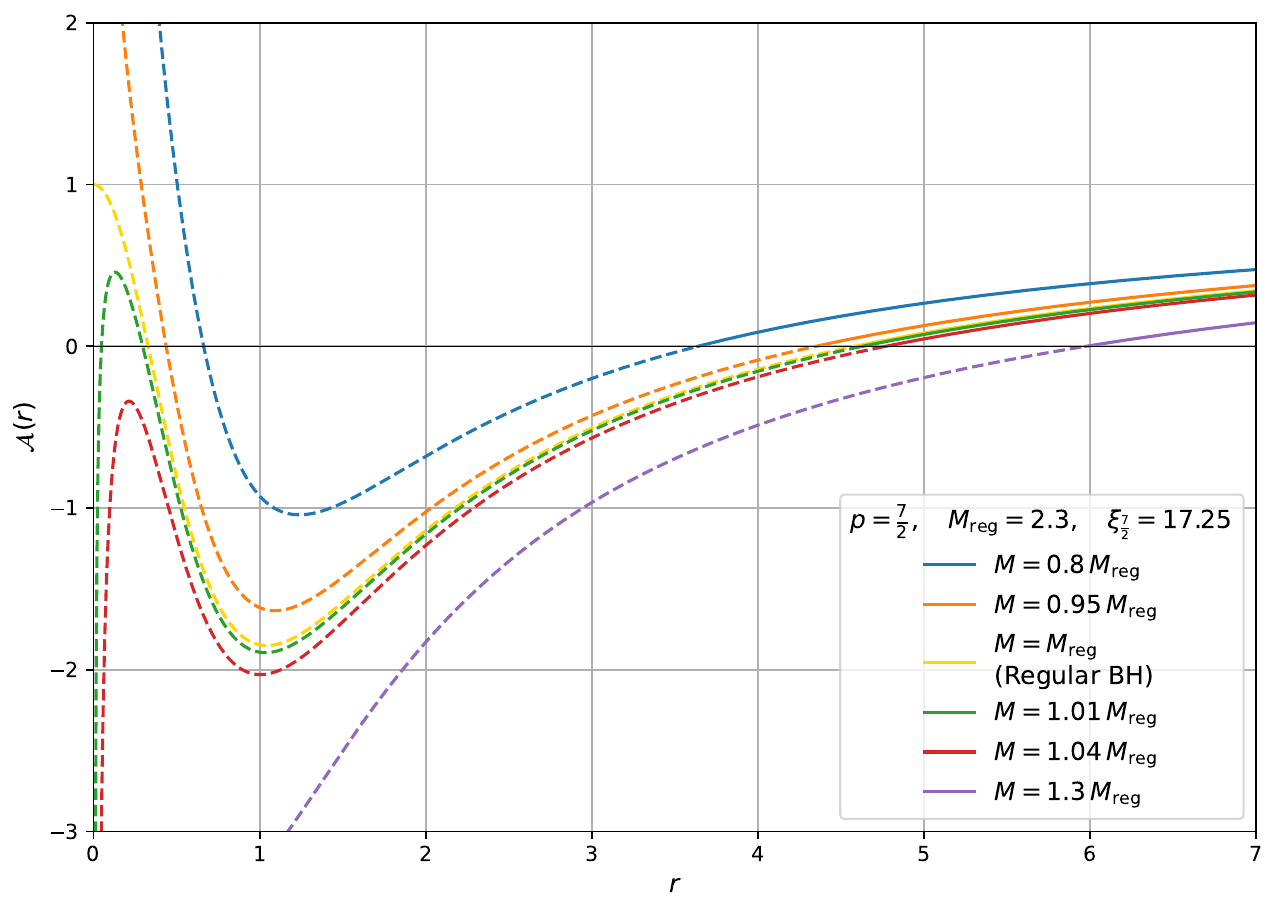}
    \end{subfigure}
    \begin{subfigure}[t]{0.49\textwidth}
        \centering
        \includegraphics[width=\textwidth]{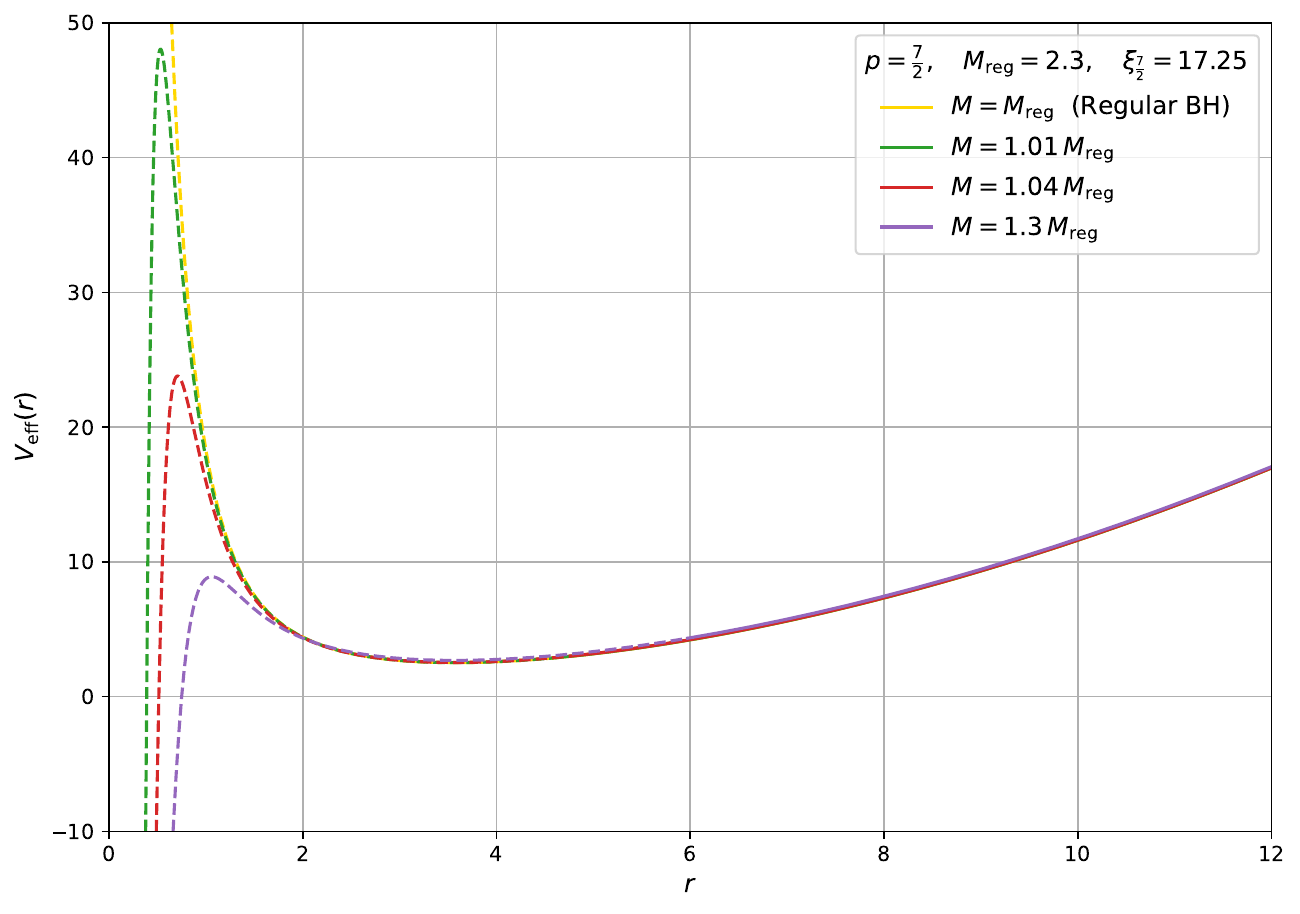}
    \end{subfigure}
    \caption{Left column: plots of the background function $\cA$ for, from top to bottom, $\nth=2$, $5/2$, and $7/2$, for different values of $M/\mreg$. 
    Right column: plots of the effective potential $\Veff$ for the same values of $\nth$ and the values $M/\mreg$ for which the background is a BH.
    The dashed parts of the curves correspond to the interior of the background BH. The yellow line corresponds to the case $\mu=M-\mreg=0$.}
    \label{fig:A_V_different_p}
\end{figure}

We have plotted the potential for different values of $\nth$ and $M/\mreg$ in the right column of Fig.~\ref{fig:A_V_different_p}.
Note that the potential \eqref{Veff_HairyBH} could be formulated in terms of $r_*$, at least formally, by inverting the relation \eqref{r_star_hairy} between $r_*$ and $r$.
In the limit $r\to 0$, one finds that the new radial coordinate goes to zero as
\begin{equation}
\label{asymrstar}
    r_* \; =\, \frac{\sqrt{2(2p-1)}}{5\sqrt{ \mu}} \,r^{5/2} (1+ {\cal O}(r))\,,
\end{equation}
while the potential diverges as\footnote{Recall that for the monopole, one had $\Veff(r_*) \; = \; 
    \frac{11 }{100\ r_*^2} (1+ {\cal O}(r_*^{4/5}))$.}
\begin{equation}
\label{V_limit_0}
    \Veff(r_*) \; = \; 
    -\frac{1 }{4 r_*^2} (1+ {\cal O}(r_*^{4/5}))\,\quad {\rm when} \quad r_*\to 0 \, ,
\end{equation}
so that, for any $\nth>3/2$, the potential diverges negatively at this end.

\noindent In the opposite limit, $r\to\infty$, $r_*$ tends to a finite limit, which we call $\rmax$,  and we have 
\begin{equation}
\label{ustar}
          u_*\, \equiv \, \rmax - r_* \; = \; \frac{\sqrt{2p-1} }{r}(1 + {\cal O}(1/r)) \,.
\end{equation}
At this end of the interval, the 
potential \eqref{Veff_HairyBH} diverges as 
\begin{equation}
\label{V_limit_max}
    \Veff\; = \, \frac{(p-2)(p-3)}{u_*^2} \left( 1 + {\cal O}(u_*)\right)\quad {\rm when} \quad r_*\to \rmax
    \qquad (p\neq2,3) \, .
\end{equation}
For these two particular values, the potential instead tends to a constant,
\begin{equation}
\Veff \to V_2^{\infty}= \frac{32 \xi_2 ^2}{27 M^2}-\frac{2}{3} \quad \text{for} \; p=2 \,,\qquad \Veff \to V_3^{\infty}= -\frac{7}{5} \quad \text{for} \; p=3 \, ,
\end{equation}
as illustrated in the right column of Fig.~\ref{fig:A_V_different_p}.

\subsection{Stability analysis}
The  construction of a self-adjoint extension of the Schr\"odinger operator (see \cite{Kral_Zettl} for instance)  depends  on the asymptotic behaviors of the solutions at both extremities of the interval $[0,\rmax]$, which can be determined explicitly. 
\subsubsection{Self-adjointness}
We start by considering the theories with $\nth >3$.
In this case, the two end points of the interval $[0,\rmax]$ are singular, since the potential is singular at the end points.
In the limit $r_*\to 0$, using \eqref{V_limit_0}, the  solutions of the Schr\"odinger equation behave as 
\begin{equation}
\label{asymptr0}
\smv=\smv_+ + \smv_- \, , \qquad 
\smv_+\left(r_*\right) = a_+ \, \sqrt{r_*}\left[1+ {\cal O}(r_*^{4/5})\right]\, , \qquad 
\smv_-\left(r_*\right) = a_- \, \sqrt{r_*}\log(r_*)\left[1+ {\cal O}(r_*^{4/5})\right]\,.
\end{equation}
Both solutions $\smv_\pm$ are square-integrable, which means that the endpoint $r_*=0$ is in the so-called limit-circle case and one must therefore impose a boundary condition here for the operator to be self-adjoint. 

On the other hand, in the limit $r_*\to \rmax$, when $p >3$ we find, using \eqref{V_limit_max},
\begin{equation}
\label{asymptr02}
\smv=\hat\smv_+ + \hat\smv_- \, , \qquad 
   \hat\smv_\pm(r_*) = b_\pm \, u_*^{s_\pm} \left[ 1 +{\cal O}(u_*) \right] \, , \quad u_*=\rmax - r_* \, ,
\end{equation}
with  $s_+=p-2$ and $s_-=3-p$. 
Only the  solution $\hat\smv_+$, which vanishes in the limit $r_*\to \rmax$, is square integrable\footnote{The second solution is square integrable only if $p<7/2$. Since $\nth$ is an integer or a half-integer, $\nth>3$ implies $\nth\geq7/2$, so that $\hat\smv_-$ is never square integrable here; note that $\nth=7/2$ is precisely the borderline case $(\nth-2)(\nth-3)=3/4$. Hence, the endpoint $r_*=\rmax$ is in the limit-point case and  there is no freedom in the boundary condition at $r_*=\rmax$.}. 
The self-adjoint extension is thus completely determined by the boundary condition imposed at the point $r_*=0$. 

There is a one-parameter  family  of boundary conditions there that lead to a self-adjoint extension of the operator. These boundary conditions can be formulated from the asymptotics \eqref{asymptr0} as follows,
\begin{eqnarray}
\label{boundcondalpha}
   a_+ \, \cos \vartheta \,  + \,  a_- \, \sin\vartheta \,  = \; 0 \, ,
\end{eqnarray}
where $\vartheta$ is an arbitrary angle at this stage. It is straightforward to check that such boundary conditions indeed lead to a symmetric operator. 
In the present context, a natural choice is the Friedrichs extension~\cite{Reed:1975mmp2}.
Although both asymptotic solutions in \eqref{asymptr0} vanish and are square-integrable near $r_*=0$, we consider for now the Friedrichs extension, which excludes the most ``irregular'' branch, i.e. the logarithmic one.  
The resulting boundary condition is therefore 
\begin{equation}
    a_-=0\, ,
\end{equation} corresponding to $\vartheta=\vartheta_F=\pi/2$ in \eqref{boundcondalpha}.

\medskip

The  cases $p=2$ and $p=3$ are special because the boundary $\rmax$ is regular\footnote{Since the potential is constant  near the boundary $\rmax$, the asymptotic solutions are  of the form 
\begin{equation}
\smv =  e^{\pm i \kappa\,  r_*} \left( 1+ {\cal O}(u_*)\right)\,, \qquad \kappa\equiv\sqrt{\omega^2-V_{2,3}^{\infty}}\,,
\end{equation}
where $u_*$ was defined in \eqref{ustar} and $\kappa$ can be imaginary (if  $\omega^2<V_{2,3}^{\infty}$), in which case the exponentials become real.  If one expands the exponential around $\rmax$, one gets the desired asymptotic solution.}.
Hence,  since $r_*=0$ is a limit-circle end point, 
there is a 2-parameter family of separated boundary conditions that lead to a self-adjoint extension.  
As in the monopole case, we also assume  for simplicity  that $\smv$ satisfies the same boundary condition at both ends as in the generic case $p >3$ discussed above, i.e.
\begin{equation}
\label{boundary_p_1_2}
  a_-=0\, \qquad \text{together with}  \qquad\smv(\rmax)=0\,, \qquad (p=2,3)\,.
\end{equation}
Such a condition implies that there is no leakage of energy (i.e. no radiation) at spatial infinity. 

Finally, there is the exotic case $p=5/2$ to consider, for which the asymptotic behavior at $r\to \infty$ is similar to the one at $r_* \to 0$:
\begin{equation}
\smv=\hat\smv_+ + \hat\smv_- \, , \qquad 
\hat\smv_+\left(r_*\right) = b_+ \, \sqrt{u_*}\left[1+ {\cal O}(u_*)\right]\, , \qquad 
\hat\smv_-\left(r_*\right) = b_- \, \sqrt{u_*}\log(u_*)\left[1+ {\cal O}(u_*)\right]\,.
\end{equation}
Both branches are square-integrable and vanish as \(u_*\to0\), so that $\rmax$ is also a limit-circle endpoint. As at the endpoint \(r_*=0\), we select the Friedrichs extension by excluding the logarithmic branch \cite{Teschl:2014}. We thus impose
\begin{equation}
b_-=0.
\end{equation}
Together with the analogous condition \(a_-=0\) at \(r_*=0\), this fixes the domain of the self-adjoint Schr\"odinger operator considered below.

To conclude, the space of square-integrable functions satisfying the chosen boundary conditions at both endpoints defines the domain of the self-adjoint Schr\"odinger operator.
We will use the notation $\SchroopF$ for this Friedrichs extension of the Schr\"odinger operator on this domain. 
The choice of boundary conditions is further justified by the next subsection on the positivity of the spectrum.

\subsubsection{Positivity of the operator}
Having defined the Friedrichs extension $\SchroopF$, we now investigate the positivity of its spectrum. If the effective potential $\Veff$ were non-negative everywhere, the positivity of the operator would follow directly. However, as shown above, $\Veff$ is generically negative in some regions of the interval, so its sign alone is not sufficient to establish stability. To overcome this difficulty, we make use of the $S$-deformation method~\cite{Ishibashi:2003ap,Kimura:2017uor,Kimura:2018eiv}.
As shown in \eqref{Veff_HairyBH}--\eqref{SandVsdef_Hairy}, the effective potential can be written in terms of a deformation function $S$ and a residual potential $\Vs$. Provided that $S$ is real and continuously differentiable on $]0,\rmax[$, the Schr\"odinger differential expression admits the factorization
\begin{eqnarray}
\label{Sdeform}
    \SchroopF \; = \; -\partial_*^2 + \Veff \; = \; \Dope^\dagger \Dope + \Vs\quad \text{with} \quad \Dope=\partial_* + S \quad \text{and} \quad \Dope^\dagger = -\partial_* + S \, .
\end{eqnarray}
One can prove the positivity of the operator after the following integration by parts 
\begin{eqnarray}
\label{psiApsi}
\int_0^{\rmax} \!\!  \mathrm{d}r_* \, \bar\smv \SchroopF \smv \; = \;  {\Bterm}_0 - {\Bterm}_{\rm max}  + \, 
\int_0^{\rmax} \!\!  \mathrm{d}r_* \vert \Dope\smv \vert^2  \, 
+\int_0^{\rmax} \!\!  \mathrm{d}r_* \Vs \vert \smv \vert^2   \, ,
\end{eqnarray}
where $\smv$ is a function in the domain of $\SchroopF$ and where the boundary terms are defined by
\begin{eqnarray}
\label{bounddefinition}
{\Bterm}_0= \lim_{r_* \rightarrow 0} \bar\smv \, \Dope\smv  \, , \qquad 
 {\Bterm}_{\rm max}=\lim_{r_* \rightarrow \rmax} \!\!\! \bar\smv \, \Dope\smv .
\end{eqnarray}
By choosing boundary conditions such that 
${\Bterm}={\Bterm}_0-{\Bterm}_{\rm max}\geq0$, one obtains a positive self-adjoint operator. 
In the cases where $\Bterm$ is not positive, our sufficient condition is not satisfied, which does not mean that the spectrum is not positive but rather shows the limitation of our method. 

Note finally that, when $r_*$ ranges over a compact interval and $\SchroopF$ is positive, the spectral theorem ensures that any solution of the wave equation \eqref{waveop} can be decomposed into normal modes of $L^2\left(\left[0, \rmax\right], d r_*\right)$.

\medskip

One first computes $\Bterm_0$ near the origin using \eqref{asymptr0} and finds for every $\nth$  
\begin{equation}
\overline{\smv}\,\Dope\smv=a_-\left(\overline{a_+}+\overline{a_-}\log r_*\right)+\frac54\left(\frac{2(2p-1)}{25}\right)^{3/5}\mu^{2/5}\,\abs{a_+}^2\,r_*^{4/5}+{\cal O}\!\left(\abs{a_-}\,r_*^{4/5}\log^2r_*\right)+{\cal O}\!\left(\abs{a_+}^2\,r_*^{8/5}\right)\, ,
\end{equation}
where the first ${\cal O}$ collects all the remaining terms proportional to $a_-$ or $\overline{a_-}$ (whose coefficients may also involve $a_+$), while the second one collects the corrections that survive when $a_-=0$. With the boundary condition $a_-=0$, this reduces to $\overline{\smv}\,\Dope\smv\propto r_*^{4/5}\abs{a_+}^2 \to0$, so that $\Bterm_0=0$.

\bigskip

Near $\rmax$, we label the two branches so that $\hat\smv_+$ always vanishes at $\rmax$:
\begin{itemize}
\item for $\nth=2$: $\hat\smv_+=b_+\,u_*\left[1+{\cal O}(u_*)\right]$ and $\hat\smv_-=b_-\left[1+{\cal O}(u_*)\right]$;
\item for $\nth=5/2$: $\hat\smv_+=b_+\sqrt{u_*}\left[1+{\cal O}(u_*)\right]$ and $\hat\smv_-=b_-\sqrt{u_*}\log u_*\left[1+{\cal O}(u_*)\right]$;
\item for $\nth\geq3$: $\hat\smv_\pm=b_\pm\,u_*^{s_\pm}\left[1+{\cal O}(u_*)\right]$, with $s_+=\nth-2$ and $s_-=3-\nth$, as in \eqref{asymptr02}; for $\nth=3$, these are the branches $u_*$ and $1$ of the regular endpoint.
\end{itemize}
The boundary condition at $\rmax$ \eqref{boundary_p_1_2} for $\nth=2,3$, the Friedrichs condition for $\nth=5/2$, and square integrability for $\nth\geq7/2$, then reads $b_-=0$ in all cases. As $r_*\to\rmax$, for a solution of the Schr\"odinger equation with frequency $\omega$,
\begin{align}
&\nth=2:\quad \overline{\smv}\,\Dope\smv=-\overline{b_-}\,b_++\frac{4\xi_2}{3\sqrt{3}\,M}\abs{b_-}^2-\abs{b_+}^2u_*+{\cal O}\!\left(\abs{b_-}\,u_*\right)+{\cal O}\!\left(\abs{b_+}^2u_*^2\right),\\
&\nth=5/2:\quad \overline{\smv}\,\Dope\smv=-b_-\left(\overline{b_+}+\overline{b_-}\log u_*\right)+\frac{\omega^2}{2}\abs{b_+}^2u_*^{2}+{\cal O}\!\left(\abs{b_-}\,u_*^2\log^2u_*\right)+{\cal O}\!\left(\abs{b_+}^2u_*^4\right),\\
&\nth\geq3:\quad \overline{\smv}\,\Dope\smv=(2p-5)\,b_-\left(\overline{b_+}+\overline{b_-}\,u_*^{5-2p}\right)+\frac{\omega^2}{2p-3}\abs{b_+}^2u_*^{2p-3}+{\cal O}\!\left(\abs{b_-}\,u_*^{7-2p}\log u_*\right)+{\cal O}\!\left(\abs{b_+}^2u_*^{2p-1}\right),
\end{align}
where here as well, the first ${\cal O}$ collects all the remaining terms proportional to $b_-$ or $\overline{b_-}$ whose coefficients may also involve $b_+$, while the second one collects the corrections that survive when $b_-=0$.
With $b_-=0$, the only remaining terms are proportional to $\abs{b_+}^2$ and vanish as $u_*\to0$, so that $\Bterm_{\rm max}=0$.

To conclude, both boundary terms therefore vanish, and \eqref{psiApsi} reduces to 
\begin{equation}
\int\bar\smv\,\SchroopF\smv\,\mathrm{d}r_*=\int\big(\abs{\Dope\smv}^2+\Vs\abs{\smv}^2\big)\,\mathrm{d}r_*>0\, 
\end{equation} 
which proves the positivity of the spectrum for $\nth\geq2$ when $\mu>0$.

\subsection{Regular BH and Soliton}
Let us now consider the special case $\mu=M-\mreg=0$, which corresponds to a regular compact object at $r=0$. The potential behaves near the origin as
\begin{eqnarray}
\label{asymptV_mu0}
    \Veff=\frac{12}{r_*^2}\left(1+{\cal O}(r_*^2)\right)\,.
\end{eqnarray}
Hence, solving the Schr\"odinger equation when $r\to0$, we get
\begin{equation}
\label{asymptu0Mu0}
\smv=\smv_+ + \smv_- \, , \qquad
\smv_\pm\left(r_*\right) = a_\pm \, r_*^{s_\pm}\left[1+ {\cal O}(r_*^2)\right]\,,
\end{equation}
with $s_+=4$ and $s_-=-3$. Since only $\smv_+$ is square integrable, $r=0$ is a limit-point endpoint, and no additional boundary condition is needed to obtain a self-adjoint operator. Then, using $S\sim 3/r_*$,
\begin{equation}
  \overline{\smv}\,\Dope\smv=7\,\abs{a_+}^2\,r_*^7+{\cal O}(r_*^9)\,,
\end{equation}
which is positive and vanishes as $r\to0$. Therefore $\Bterm_0=0$ again and, together with the analysis at $r\to\infty$, $\Bterm=0$ and the regular object is always stable under linear perturbations.

\subsection{Beyond the Friedrichs extension and instabilities}
\label{subsec:beyond_Friedrichs}
The positivity of the spectrum established above holds for the Friedrichs extension $\SchroopF$.
For singular configurations ($\mu\neq0$), the endpoint $r_*=0$ is of limit-circle type, and the Schr\"odinger operator admits other self-adjoint realizations, distinguished by the boundary condition imposed at the singularity. As can be seen from eq.~\eqref{boundcondalpha}, these boundary conditions are labelled by an angle $\vartheta$, defined modulo $\pi$, so that they form a circle, on which the Friedrichs extension corresponds to $\vartheta_F=\pi/2$. A similar one-parameter family of boundary conditions at a timelike singularity was studied for the negative-mass Schwarzschild spacetime in Ref.~\cite{Gibbons_2005}, where some of them were shown to lead to stable dynamics and others to instabilities. It is thus natural to ask which of these extensions have a non-negative spectrum.
By contrast, for regular configurations ($\mu=0$), the origin $r=0$ is a limit-point endpoint, leaving no freedom in the choice of boundary condition there, so that, with the boundary condition at the other endpoint fixed as above, the self-adjoint realization is unique and has a positive spectrum.
We now investigate the other self-adjoint extensions available for singular configurations, first for the monopole, for which this question was raised in Appendix~D of Ref.~\cite{Iteanu2026rph}, and then for the dipole.

We will show that, for both sectors, the transition between positive and negative spectra is governed by the zero mode $\zeromode$ satisfying
\begin{equation}
\label{eq:Static_solution}
  \Schroop\zeromode=0,
  \qquad \text{with the prescribed boundary condition at }r_*=\rmax.
\end{equation}
Its behaviour near the origin selects a critical angle $\vartheta_c$. The two points $\vartheta_F$ and $\vartheta_c$ divide the circle of boundary conditions into two arcs. On the arc going from $\vartheta_F$ to $\vartheta_c$ in one direction, the spectrum is positive; at $\vartheta=\vartheta_c$, the lowest eigenvalue vanishes; and on the complementary arc, there is exactly one negative eigenvalue, which tends to $-\infty$ as $\vartheta$ approaches $\vartheta_F$.
In other words, the system remains stable when one moves away from the Friedrichs boundary condition in one direction, up to the critical one, whereas it is unstable for all the other boundary conditions (see Fig.~5 of Ref.~\cite{Gibbons_2005} for an illustration).
Here, we prove this analytically using some standard tools of Sturm-Liouville analysis: self-adjoint boundary conditions for singular Sturm-Liouville operators~\cite{Gesztesy:2019} (Theorems~3.10 and 3.11, and Remark~3.12 when $\rmax$ is limit point), together with the min-max principle~\cite{Teschl:2014} (Theorems~4.12 and 4.14 and Corollary~4.13).
The resulting comparison of eigenvalues for different self-adjoint extensions is closely related to the results of Ref.~\cite{Niessen:1992}.

\subsubsection{Generic boundary conditions}
\label{subsubsec:bc_circle}

We keep the boundary condition at $\rmax$ chosen in 
Sec.~\ref{Sec:HairyBH} (and in Ref.~\cite{Iteanu2026rph} for the monopole), but let the condition at the origin vary. 
For the dipole, it is parametrised by the angle $\vartheta$ in \eqref{boundcondalpha}. 
For the monopole, the master variable $\Psi$ behaves near the origin as
\begin{equation}
\label{bc_theta_monopole}
\Psi\simeq c_+\,r^{11/4}\left(1+{\cal O}(r^2)\right)+c_-\,r^{-1/4}\left(1+{\cal O}(r^2)\right),
\qquad c_+\cos\theta+c_-\sin\theta=0,
\end{equation}
where the two branches are normalised as in Eq.~(3.38) of Ref.~\cite{Iteanu2026rph}.
In terms of $r_*$, these two branches behave as $r_*^{11/10}$ and $r_*^{-1/10}$.
We denote the self-adjoint Schr\"odinger operators by $\Schroop_\theta$ for the monopole and $\Schroop_\vartheta$ for the dipole and in the following, $\Schroop_\eta$ denotes either $\Schroop_\theta$ or $\Schroop_\vartheta$, with $\eta=\theta$ or $\eta=\vartheta$, and $\SchroopF$ denotes the Friedrichs extension in the corresponding sector.
Within each sector, fixing the boundary condition at $\rmax$ leaves only the limit-circle endpoint at the origin free. The self-adjoint extensions of the resulting symmetric operator form a one-parameter family, see Theorem~9.6 and the discussion following in Ref.~\cite{Teschl:2014}.

Only the ratio of the two asymptotic coefficients matters, so the angles are defined modulo $\pi$. Thus $\pi/2$ and $-\pi/2$ represent the same point on the circle of boundary conditions, namely the Friedrichs extension ($c_-=0$ for the monopole, $a_-=0$ for the dipole), as depicted in Fig.~5 of Ref.~\cite{Gibbons_2005}.

Standard results for singular Sturm-Liouville operators give two useful properties of these families.
First, all these operators have a discrete, simple spectrum as proven in Sec.~9.2 of Ref.~\cite{Teschl:2014}. 
Denoting by $\omega_0^2(\eta)<\omega_1^2(\eta)<\cdots$ their eigenvalues and by $\omega_{n,\mathrm F}^2>0$ those of the Friedrichs extension, one has the strict interlacing inequalities from Ref.~\cite{Niessen:1992} and the min-max principle (Theorem~4.14 of~\cite{Teschl:2014})
\begin{equation}
\label{interlacing}
  \omega_{n-1,\mathrm F}^2<\omega_n^2(\eta)<\omega_{n,\mathrm F}^2,
  \qquad \eta\neq\pm\frac{\pi}{2},
\end{equation}
with $\omega_{-1,\mathrm F}^2\equiv-\infty$ by convention. In particular,
$\omega_1^2(\eta)>\omega_{0,\mathrm F}^2>0$: irrespective of the boundary condition, there can be at most one negative eigenvalue, and it must be the fundamental one ($n=0$).

Second, the eigenvalues vary monotonically with the boundary angle. Differentiating the eigenvalue equation with respect to the angle and applying Green's formula (Eq.~(9.7) of Ref.~\cite{Teschl:2014}) reduces the derivative to the boundary Wronskian at the origin, $\mathcal W[f,g]=f\,\partial_*g-g\,\partial_*f$. Evaluating it on the asymptotic branches, $\mathcal W[r^{11/4},r^{-1/4}]=-3{\cal K}$ for the monopole and $\mathcal W[\sqrt{r_*},\sqrt{r_*}\log r_*]=1$ for the dipole, gives
\begin{equation}
\label{monotonicity}
\begin{aligned}
 \frac{\partial\omega_n^2}{\partial\theta}
 =\frac{-3{\cal K}\,c_-^2}
 {\cos^2\theta\int_0^{\rmax}\Psi_n^2\,\mathrm{d}r_*},\qquad 
 \frac{\partial\omega_n^2}{\partial\vartheta}
 =\frac{a_-^2}
 {\cos^2\vartheta\int_0^{\rmax}\smv_n^2\,\mathrm{d}r_*},
\end{aligned}
\end{equation}
where ${\cal K}=\sqrt{2\mu/(2p-1)}$ and $c_-$ and $a_-$ are the respective asymptotic coefficients of the eigenfunctions. Hence, on $]-\pi/2,\pi/2[$, the eigenvalues decrease strictly with $\theta$ for the monopole and increase strictly with $\vartheta$ for the dipole.

Together, interlacing and monotonicity imply that only the lowest
eigenvalue can cross zero, and that it can do so at most once as the
angle varies over $]-\pi/2,\pi/2[$. The critical boundary condition is
therefore determined by the static solution of \eqref{eq:Static_solution}
that satisfies the prescribed condition at $\rmax$: the critical angle
is the one for which this same solution also satisfies the condition
at the origin.

\subsubsection{The monopole}
\label{subsubsec:monopole_threshold}

For the monopole, the deformation function $\smono$, given by \eqref{Sfunction_Monopole_Hairy}, depends on the integration constant $c_S$. It is regular on $]0,\rmax[$ as long as $\Xi_p(r)+c_S$ does not vanish, i.e.\ for $c_S\notin\,]-\mreg/\xi_p,0[$. For these values, the boundary term $\Bterm_0$ in \eqref{psiApsi} is non-negative and the spectrum of $\SchroopEt$ is positive provided (Appendix~D of Ref.~\cite{Iteanu2026rph})
\begin{equation}
\label{D6_monopole}
  \tan\theta< -\frac13\left(\frac1{c_S}
              -\frac{7(2\xi_p+3)}{8\mu}\right).
\end{equation}
Since $c_S$ enters only through the factorisation, the operator and its spectrum do not depend on it; each admissible value of $c_S$ provides an independent sufficient condition for positivity. The strongest bound is obtained at the edge $c_S=-\mreg/\xi_p$, giving the critical
angle $\theta_c$ defined by
\begin{equation}
\label{thetac_monopole}
  \tan\theta_c=\frac{\xi_p}{3\mreg}
                    +\frac{7(3+2\xi_p)}{24\mu}.
\end{equation}
This proves positivity for $-\pi/2\leq\theta<\theta_c$. We now show that the bound is also the exact threshold for instability.

The static solutions of $\Schroop f=0$ form a two-dimensional space, spanned by $u_0=(1+r^2)^{p/2}/(r\sqrt{n})$ and $u_0\,\Xi_p$: up to normalisation, they are $u_0\,(\Xi_p+c_S)$ with $c_S\in\mathbb R$, together with $u_0$ itself, obtained in the limit $c_S\to\infty$. For $c_S\notin\,]-\mreg/\xi_p,0[$, where $\smono$ is regular, these are precisely the zero modes $\exp(-\int\smono\,\mathrm{d}r_*)$ of $\Dope$.

For generic $c_S$, the zero mode behaves near $\rmax$
as $u_*^{2-p}$, which does not satisfy the imposed boundary condition
(and is not square-integrable for $p\geq5/2$). For
$c_S=-\mreg/\xi_p$, it behaves as the admissible branch $u_*^{p-1}$
(see Eq.~(3.27) of Ref.~\cite{Iteanu2026rph}). Therefore
\begin{equation}
\label{zeromode_monopole}
  \zeromodeMono=\exp\!\left(-\int\smono\,\mathrm{d}r_*\right)
                 \Big|_{c_S=-\mreg/\xi_p}
\end{equation}
is, up to normalisation, the unique static solution satisfying the condition at $\rmax$. Its asymptotic coefficients at the origin obey \eqref{bc_theta_monopole} precisely when $\theta=\theta_c$.
Consequently, $\zeromodeMono$ is an eigenfunction of $\Schroop_{\theta_c}$ with eigenvalue zero. By \eqref{interlacing}, this must be the ground state: $\omega_0^2(\theta_c)=0$. Thus the sufficient positivity condition \eqref{D6_monopole} becomes exact when the deformation zero mode satisfies the boundary condition at infinity.

At the critical angle $\theta=\theta_c$, the time-dependent wave equation also admits $\Psi=(A+B t_*)\zeromodeMono$, which grows linearly for $B\neq0$;
we therefore refer to this case as marginally stable. By the strict
monotonicity \eqref{monotonicity}, $\omega_0^2$ becomes negative for
$\theta>\theta_c$. Interlacing ensures that there is no other
negative eigenvalue. In summary, the monopole has a positive spectrum
for $-\pi/2\leq\theta<\theta_c$, a zero fundamental eigenvalue at
$\theta=\theta_c$, and exactly one exponentially growing mode
$e^{\kappa t_*}$, with $\kappa^2=-\omega_0^2$, for
$\theta_c<\theta<\pi/2$.

\subsubsection{The dipole}
\label{subsubsec:dipole_threshold}

For the monopole, the critical angle was obtained explicitly from the zero mode associated with the $S$-deformation with $\Vs=0$ given in \eqref{Sfunction_Monopole_Hairy}.
For the dipole, no such closed-form solution is available. The explicit deformation \eqref{SandVsdef_Hairy} has $\Vs\neq0$
and therefore does not directly provide a static solution. Moreover, for $p\geq5/2$, this deformation cannot be tuned as in the monopole case while keeping $\Vs$ fixed
\footnote{Writing $f_S=\exp(-\int S\,\mathrm{d}r_*)$, the other solutions of the same Riccati equation are $\widetilde S=-\partial_*\ln\left| f_S(a+b\int f_S^{-2}\,\mathrm{d}r_*)\right|$. For $p\geq5/2$, the integral diverges at both endpoints: logarithmically at $r_*=0$, and logarithmically for $p=5/2$ or as a power law for $p>5/2$ at $\rmax$. Since it is strictly increasing and spans the real line, any $b\neq0$ produces a zero inside the interval and hence a singular $\widetilde S$. Thus, varying this deformation while keeping $\Vs$ fixed does not yield an analogue of \eqref{D6_monopole}.}.

Nevertheless, the critical angle is still determined by the static solution $\zeromode$ satisfying the boundary condition at $\rmax$. This solution can be obtained numerically by integrating $\Schroop\zeromode=0$ inward from its asymptotic behaviour at $\rmax$. Near the origin, it behaves as
\begin{equation}
\label{zeromode_origin}
\zeromode\simeq
a_+^{(0)}\sqrt{r_*}
+a_-^{(0)}\sqrt{r_*}\log r_*,
\end{equation}
so that the critical angle $\vartheta_c$ is defined by
\begin{equation}
\label{thetac_def}
\tan\vartheta_c=-\frac{a_+^{(0)}}{a_-^{(0)}}.
\end{equation}
Here $a_-^{(0)}\neq0$, since the Friedrichs extension has a strictly positive spectrum.

At $\vartheta=\vartheta_c$, the static solution is a zero eigenfunction, necessarily the ground state by \eqref{interlacing}. Together with the monotonicity \eqref{monotonicity}, this implies that the dipole has a positive spectrum for $\vartheta_c<\vartheta\leq\pi/2$, is marginally stable at $\vartheta=\vartheta_c$, and has exactly one unstable mode for $-\pi/2<\vartheta<\vartheta_c$.

\begin{figure}[t]
\centering
\includegraphics[width=\textwidth]{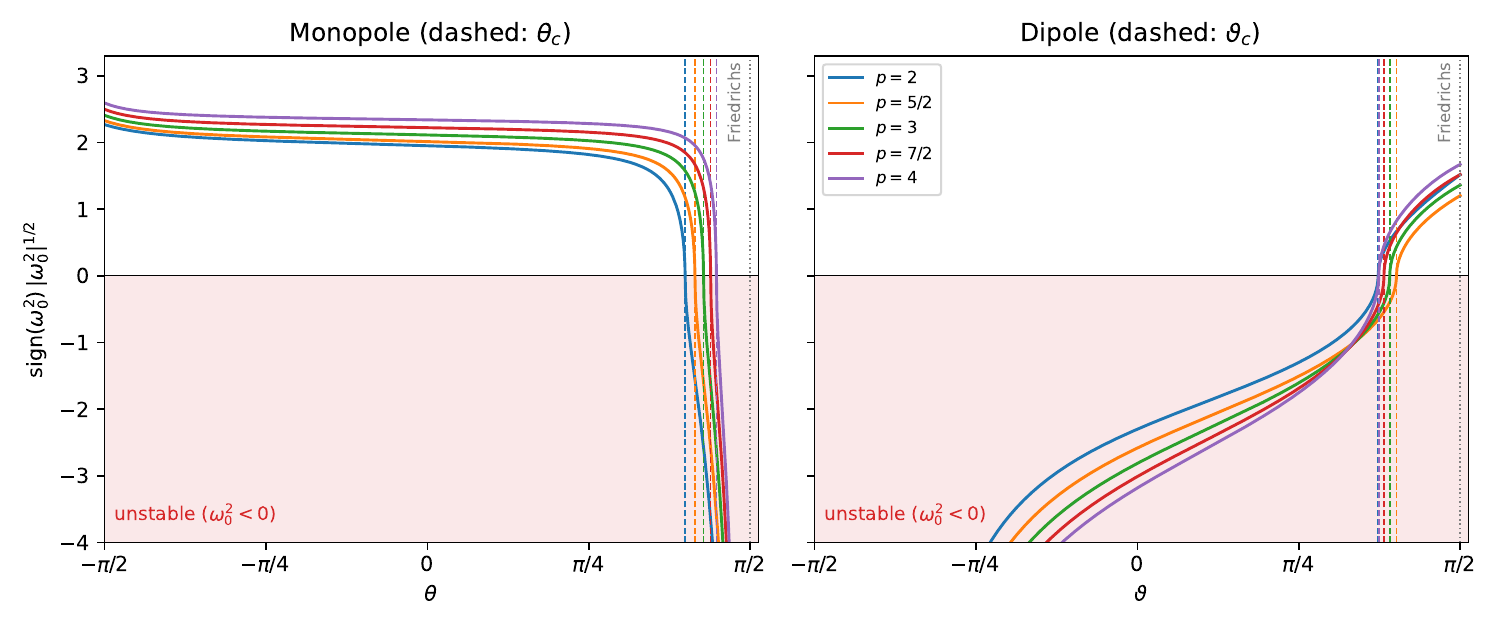}
\caption{Lowest eigenvalue of the Schr\"odinger operator as a function of  the boundary angle at the origin, for the monopole (left) and the dipole  (right), with $\mu=1$ and $\xi_p=3$, for $\nth=2,5/2,3,7/2,4$. The quantity shown is  $\mathrm{sign}(\omega_0^2)\,|\omega_0^2|^{1/2}$, i.e.\ the frequency of  the fundamental mode for stable extensions and minus its growth rate  for unstable ones. The shaded region corresponds to the unstable region $\omega_0^2<0$,  the dashed lines to the critical angles $\theta_c$ and $\vartheta_c$,  and the dotted line to the Friedrichs extension ($\pm\pi/2$).}
\label{fig:fundamental_vs_angle}
\end{figure}

We have confirmed these results by solving numerically the eigenvalue problem $\Schroop\smv=\omega^2\smv$ for several hundred values of the boundary angle, imposing the corresponding asymptotic conditions at the origin \eqref{bc_theta_monopole} and \eqref{asymptr0}.
The results are shown in Fig.~\ref{fig:fundamental_vs_angle}.
For the monopole, the angle at which $\omega_0^2$ vanishes agrees with the analytical prediction \eqref{thetac_monopole}.
For the dipole, it agrees with the critical angle \eqref{thetac_def} obtained by independently integrating the static equation. In both cases, the agreement is at the level of $10^{-6}$ in relative accuracy.

\section{Non-propagating scalar in Horndeski theory}
\label{section:Non_propagating}
\subsection{The $\nth=1/2$ theory}
\label{subsection_p=1/2}
The family of theories \eqref{P_n} contains a special case, characterized by $\nth = 1/2$, which belongs to the subclass of Horndeski theories:
\begin{equation}
P(X)=-2 \alpha \sqrt{X}, \quad F(X)=1-\alpha \sqrt{X}, \qquad A_1(X)=-F_X=\frac{\alpha}{2 \sqrt{X}} \, .
    \label{n=1/2}
\end{equation}
This theory admits a stealth solution \cite{Bakopoulos:2023fmv} for which the metric coincides with the Schwarzschild solution while supporting a non-trivial scalar field  
\begin{eqnarray}
\label{Schwarmetric}
    \cA(r)=\cB(r)=1-\frac{2M}{r} \, , \qquad X(r)=\frac{  q^2}{2\left(1+r^2\right)} \, .
\end{eqnarray}

Let us recall here that the
determinant $\Det$ entering the generic monopole reduction of Ref.~\cite{Iteanu2026rph} vanishes for this theory, leading to a non-propagating degree of freedom at the linear level. This is confirmed by the following dipole analysis. Indeed, one can follow the procedure described in Sec.~\ref{perturbations section} and obtain a second-order equation of the type \eqref{def_second_order_equation} in the original coordinate system $(t,r)$:
\begin{equation}
\begin{aligned}
&{\psi'(r)^2} \frac{\partial^2 \mv}{\partial t^2} + \frac{q^2}{\nu(r)}   \frac{\partial}{\partial r} \left( \nu(r)\frac{\partial \mv}{\partial r} \right)- {2q \psi'(r)}  \frac{\partial^2 \mv}{\partial t \partial r} 
\;+
\frac{q^2 r^2 \sqrt{r^2+1} \left(-\left(6 r^2+1\right) \cA+12 r^4+10 r^2+1\right) }{\nu(r)}\mv\\
&+\frac{q^3  \left(2 \left(r^2+1\right)^2-\left(20 r^2+9\right) \left(r^2+1\right) \cA+\left(11 r^2+7\right) \cA^2\right)}{2 r \left(r^2+1\right)^2 \cA^3 \psi '(r)}\dot{\mv} = \; 0 \, ,
\label{Second_order_p12_tr}
\end{aligned}
\end{equation}
where we have defined
\begin{eqnarray}
    \nu(r)=r^4\left(r^2+1\right)^{3/2} \left(1 +   r^2 -  \cA \right) \, .
\end{eqnarray}
Eq.~\eqref{Second_order_p12_tr} looks like a wave equation, as it contains second-order time derivatives.
However, one can reabsorb all the time derivative terms by redefining the time coordinate through
\begin{eqnarray}
\label{unitary}
    t_*=\phiback/q \, , \qquad\text{such that the prime now denotes}\qquad
   \left.\frac{\partial}{\partial r}\right|_{t_*}  =  \left.\frac{\partial}{\partial r}\right|_t- \frac{\psi'}{q} \frac{\partial}{\partial {t_*}} \,.
\end{eqnarray}
Then, in this new system of coordinates, the perturbation equation simplifies drastically and reduces to 
\begin{equation}
\mv''
+3 \left(\frac{r}{r^2+1-\cA}+\frac{2r}{r^2+1}+\frac{1}{r^3+r}\right) \mv^{\prime}
+\frac{\left(12 r^4+10 r^2+1-\left(6 r^2+1\right) \cA\right) }{r^2 \left(r^2+1\right) \left(r^2+1-\cA\right)}\mv=0\,,
\label{SecondOrderN12}
\end{equation}
where no derivative with respect to $t_*$ appears.
This equation can be integrated, and the general solution for $\mv$ depends on two arbitrary functions $f(t_*)$ and $g(t_*)$ that should be fixed by appropriate (boundary or initial) conditions. Hence, the dipole is not propagating and behaves as a shadowy (instantaneous) mode \cite{DeFelice:2018ewo} as explained hereafter.
As already noted in \cite{Sirera:2024ghv}, this is linked to the fact that this theory belongs to the class of `extended cuscuton' theories \cite{Iyonaga_2018}. 
Indeed, in ADM form \cite{Langlois:2015skt}, 
\begin{equation}
 \mathrm{d}s^2 = -N^2\,dt_*^2 + \gamma_{ij}\,(dx^i + N^i dt_*)(dx^j + N^j dt_*) ,
  \qquad \sqrt{-g} = N\sqrt{\gamma} ,
\end{equation}
so that 
\begin{eqnarray*}
    X = q^2/(2N^2)\, .
\end{eqnarray*} 
As demonstrated in~\cite{Gleyzes:2014dya}, in unitary gauge, the general quadratic Horndeski Lagrangian reads
\begin{equation}
  \mathcal{L} = N\sqrt{\gamma}\,\Big[\,P(X) + F(X)\,R^{(3)} + \big(F(X) + 2XA_{1}(X)\big)\big(K_{ij}K^{ij} - K^2\big)\Big] ,
\end{equation}
where for the theory \eqref{n=1/2},
\begin{equation}
  F + 2XA_{1} = 1\, ,
  \qquad
  N\sqrt{\gamma}\,\sqrt{X} = \frac{q}{\sqrt 2}\,\sqrt{\gamma} ,
\end{equation}
so the kinetic term is exactly that of GR and the lapse drops out of all the other terms. The action becomes
\begin{equation}
  S = \int d^4x\, N\sqrt{\gamma}\,\big[R^{(3)} + K_{ij}K^{ij} - K^2\big]
      \;-\; \frac{\alpha q}{\sqrt{2}} \int d^4x\,\sqrt{\gamma}\,\Big(R^{(3)} + 2\Big) .
  \label{eq:S_unitary_p12}
\end{equation}
The first term is the GR action. The second depends only on the spatial metric $\gamma_{ij}$ of the $\phi=\text{const}$ slices: neither the lapse, nor the shift, nor any time derivative appears. This is the mechanism of the original cuscuton, for which $N\sqrt{\gamma}\sqrt{X}\propto\sqrt{\gamma}$.

Let us also mention one very interesting feature of this theory. If one considers the linear perturbations of the scalar field about a fixed background (i.e. taking $h_{\mu \nu}=0$ in \eqref{pertgeneral}), one obtains a second-order equation for $\delta \phi$ only. If one then renormalizes the scalar perturbation, one obtains exactly the same equation \eqref{SecondOrderN12} in the same coordinate $t_*$.

\subsection{4D Einstein-Gauss-Bonnet BH}
\label{subsec:4DEGB}
In \cite{Iteanu2026rph}, we had found that, for the BH solution in 4D Einstein-Gauss-Bonnet theory \cite{Charmousis:2021npl}, $\Det=0$. 
There is, however, an important difference with the $p=1/2$ theory. For the latter, the instantaneous mode is a property of the theory itself: in unitary gauge, the extra term coming from the scalar sector in the  action is independent of the lapse, so that the scalar does not propagate on any background where $\partial_\mu\phi$ is timelike, and the instantaneous slices are the $\phi=\text{const}$ hypersurfaces. In the 4D Einstein-Gauss-Bonnet case, the slicing in which the mode becomes instantaneous (i.e. where the second-order equation for the dipole does not contain any time derivative) is instead
\begin{equation}
  t_* = t + \int \frac{q\, r}{\cA\sqrt{q^2 r^2 + \cA}}\, dr ,
\end{equation}
which does not coincide with the unitary-gauge slicing $\phi=\text{const}$ of the shadowy mode.
Moreover, this theory is not of the cuscuton type: on a cosmological background, the unitary-gauge Hamiltonian is not linear in the lapse, and the scalar propagates as can be seen using the formulas of \cite{Kobayashi_2011}.
The absence of a propagating scalar mode is therefore a property of this particular family of black hole solutions, not of the theory. It signals that the kinetic term of the scalar perturbation vanishes on this background, which suggests that the scalar is strongly coupled there, rather than that the theory propagates only the tensor degrees of freedom.

Nonetheless, we have explicitly checked that the previous remark also applies to this solution: up to a renormalization of the field, the equation for the dipole is identical to the one for the scalar field about a fixed background (i.e. taking $h_{\mu \nu}=0$).

\section{Stealth solution with a constant kinetic density $X$}
\label{stealth section}
In this section, we study the dipole in another class of stealth solutions in Horndeski theories, where the kinetic density $X$ is constant.  

\subsection{Background theory and equations}
A class of Horndeski theories with a vanishing stress-energy tensor, and thus admitting stealth solutions, can be described by
\begin{eqnarray}
\label{stealth_X0_conditions}
    X=X_0= q^2/2, \quad P\left(X_0\right)=P_X\left(X_0\right)=Q_X\left(X_0\right)=0 \, .
\end{eqnarray}
Since $X$ is a constant $X_0$, the scalar field satisfies
\begin{eqnarray}
\label{stealthfiedl}
    \phi=qt+\psi(r) \qquad \text{with} \qquad \psi'(r)^2 = \frac{{ 2q^2Mr}}{(r-2M)^2} \, .
\end{eqnarray}
Furthermore, one can restrict, without loss of generality, the analysis of linear perturbations to a subclass of Horndeski theories defined by the functions
\begin{equation}
\begin{aligned}
& F(X) \equiv 1+f_1\left(q^2-2X\right)+\frac{f_2}{2}\left(q^2-2X\right)^2 \,,\qquad  P(X) \equiv \frac{p_2}{2}\left(q^2-2X\right)^2
\, ,
\end{aligned}
\end{equation}
where  $f_1$, $f_2$, $p_2$ are constants (note that we changed the name of the coupling constant with respect to \cite{Iteanu2026rph}).
Indeed, these functions satisfy the above conditions \eqref{stealth_X0_conditions},  and we have verified that any higher-order term  in powers of $\left(q^2-2X\right)$ does not contribute to the linearized equations. Note also that the previous parameter $q$ now appears in the theory functions and is thus fixed. 

\subsection{Dipole equations}

Although the dipole in such theories has already been studied in the literature \cite{deRham:2019gha,Khoury:2020aya,Takahashi:2021bml}, it is instructive to revisit it with our approach, as we also provide a stability analysis.

We first construct the coordinate system $(t_*,r_*)$ in which the second-order equation for the master variable takes a Schr\"odinger-like form. As for the hairy BH solutions, the Schr\"odinger coordinates governing the dipole coincide with those obtained for the monopole. We introduce the length scale $\reff$, defined by
\begin{equation}
    \reff^2=4\frac{(f_1^2+f_2)(4f_1q^2-4f_2q^4+1)}{p_2(2f_1q^2+1)^2}\,,
\end{equation}
where we assume $p_2\neq0$, $f_1^2+f_2\neq0$ and $2f_1q^2+1\neq0$. We further require $\reff^2>0$ for the second-order equation to be hyperbolic, as discussed in \cite{Iteanu2026rph}, and take $\reff>0$.

It is convenient to introduce  the dimensionless radial coordinate
\begin{equation}
    x=\frac{r}{\reff}\,,
\end{equation}
together with the Schwarzschild radius $\rsc=2M$. The function $n$, which defines the radial Schr\"odinger coordinate through \eqref{def_rstar}, then reads
\begin{equation}
\label{nstealth}
    n(r)=\frac{\rsc^{1/2}}{\reff^{1/2}}\,\nu(x)\,,
    \qquad
    \nu(x)=\frac{x^2-1}{x^{3/2}}\,.
\end{equation}

We restrict ourselves to the region $r>\reff$, or equivalently $x>1$, where $n>0$. Introducing the dimensionless Schr\"odinger coordinate $x_*$, we obtain
\begin{equation}
\label{xstar_stealth}
    r_*=\frac{\reff^{3/2}}{\rsc^{1/2}}\,x_*\,,
    \qquad
    \mathrm{d}x_*=\frac{\mathrm{d}x}{\nu(x)}\,,
\end{equation}
which can be integrated explicitly to give, up to an irrelevant additive constant,
\begin{equation}
    x_*(x)=2\sqrt{x}
    +\frac12\ln\left(\frac{\sqrt{x}-1}{\sqrt{x}+1}\right)
    -\arctan\sqrt{x}\,.
\end{equation}
Similarly to the usual tortoise coordinate in GR, $x_*$ (and therefore $r_*$) ranges over the whole real line, with $x_*\to-\infty$ as $x\to1^+$ and $x_*\to+\infty$ as $x\to+\infty$.

Next, the time coordinate $t_*$ is constructed through \eqref{def_tstar}, using
\begin{equation}
    W(r)=\frac{\psi'(r)}{q}
    +\frac{r^{5/2}}{\rsc^{1/2}(r^2-\reff^2)}\,.
\end{equation}
This expression depends on the choice of branch for the background scalar field \eqref{stealthfiedl}. We choose the positive branch,
\begin{equation}
    \psi(r)=q\int\frac{\sqrt{1-\cA(r)}}{\cA(r)}\,\mathrm{d}r
    =q\left[
    2\sqrt{\rsc r}
    -\rsc\ln\left|
    \frac{\sqrt{r}+\sqrt{\rsc}}{\sqrt{r}-\sqrt{\rsc}}
    \right|
    \right]+\mathrm{const.}\,,
\end{equation}
for which the time coordinate $t_*$ is regular at the Schwarzschild horizon (see \cite{Iteanu2026rph} for a proof).

In terms of the dimensionless coordinate $x_*$, the Schr\"odinger-like equation for the dipole takes the form
\begin{equation}
\label{Schro_stealth_x}
    -\frac{\mathrm{d}^2\smv}{\mathrm{d}x_*^2}
    +V(x)\smv
    =\Omega_0^2\smv\,,
\end{equation}
where the dimensionless frequency and potential are defined by
\begin{equation}
    \Omega_0=\frac{\reff^{3/2}}{\rsc^{1/2}}\omega\,,
    \qquad
    \Veff(r)=\frac{\rsc}{\reff^3}V(x)\,.
\end{equation}
The effective potential is explicitly given by
\begin{equation}
\label{Potentiel_Stealth}
    V(x)=-\frac{(x^2-1)(x^2-25)}{16x^5}\,.
\end{equation}
We now have all the ingredients entering the Schr\"odinger-like equation for the dipole.

\subsection{Stability analysis}

To establish the stability of the dipole, we follow the same approach as in the previous sections and apply the $S$-deformation method. 
Since $V$ is real and bounded and $x_*$ ranges over the whole real line, both endpoints are in the limit-point case. The Schr\"odinger operator $\Schroop$ admits a unique self-adjoint realization $\Schroop_{SA}$ in $L^2(\mathbb R,\mathrm{d}x_*)$. Thus, no boundary condition needs to be specified. Note also that $V$ is negative at large $x$, $V\simeq -1/(16x)$, so that the non-negativity of the operator is not obvious a priori.

To prove this property, we observe that the dimensionless potential \eqref{Potentiel_Stealth} admits the decomposition
\begin{equation}
\label{Veff_Stealth}
    V(x)=S^2-\frac{\mathrm{d}S}{\mathrm{d}x_*}
    +\Vs(x)\,,
\end{equation}
where
\begin{equation}
\label{SandVsdef_Stealth}
    S(x)=-\frac{\mathrm{d}}{\mathrm{d}x_*}\ln x^{1/4}
    =-\frac{x^2-1}{4x^{5/2}}\,,
    \qquad
    \Vs(x)=\frac{x^2-1}{x^5}\,.
\end{equation}
Since $S$ is smooth and bounded on $]1,+\infty[$, and vanishes at both endpoints, the boundary term in the $S$-deformation identity \eqref{psiApsi} vanishes for any function in the domain $H^2(\mathbb R)$ of the self-adjoint Schr\"odinger operator. Moreover, $\Vs>0$ throughout this interval, so the operator is non-negative, proving the mode stability of the dipole.

\section{Conclusion}
We have investigated the dipolar oscillations of black hole solutions within the framework of DHOST theories. Our main objective was to further assess the dipolar stability of several static black hole solutions reported in the literature, particularly the recently discovered solutions with primary hair \cite{Bakopoulos:2023fmv,Baake:2023zsq,Bakopoulos:2023sdm, Charmousis:2025xug}, but earlier solutions too \cite{Minamitsuji:2019shy, Charmousis:2021npl}.

We have first derived the equation of motion for a single physical degree of freedom --- a gauge-invariant quantity mixing scalar and metric perturbations. The whole computation can be found on GitHub \cite{Iteanu2026Git}.
We have then shown that the time and radial coordinates used to recast it in a Schr\"odinger-like form (or equivalently the effective metric)  are equal to those of the monopole in general.

We applied our procedure to the black hole solutions with primary hair \cite{Bakopoulos:2023fmv,Baake:2023zsq,Bakopoulos:2023sdm}, which represent a one-parameter deformation of the Schwarzschild BH parametrised by the scalar charge. As seen in \cite{Iteanu2026rph}, for this particular family of solutions, the ``Schr\"odinger'' time coordinate coincides with the scalar field itself and the new ``Schr\"odinger'' radial coordinate covers the whole spacetime,  from spatial infinity down to the BH central singularity, while spanning a bounded interval as for the monopole.
The two perturbations thus live within the same effective metric.
We also explicitly checked that the Schr\"odinger-like equation obtained for one solution remains the same for all solutions that are connected  to the original metric via invertible conformal-disformal transformations, as was the case in \cite{Iteanu2026rph}.

The resulting Schr\"odinger-like equation, defined on a finite interval, is characterised by its potential, which always contains a negative part. Moreover, we found that this potential diverges at both ends of the interval (i.e.\ at the singularity and at spatial infinity), except for the theories $\nth=2$ and $\nth=3$, where it remains finite at spatial infinity.
We studied the boundary conditions for these singular Sturm-Liouville problems and showed, using an $S$-deformation, that the Friedrichs extension is a positive self-adjoint operator, which ensures the dipolar stability of the solution. 
Moreover, since the singularity is a limit-circle endpoint, we also explored different self-adjoint extensions parametrized by an angle $\vartheta$.
We showed that not all of them are stable.
Indeed, the static solution which satisfies the boundary condition at spatial infinity selects a critical angle $\vartheta_c$, at which it is a zero mode, so that this extension is only marginally stable. We showed that extensions lying on the same side of $\vartheta_c$ as the Friedrichs one have a positive spectrum, while all the others have exactly one exponentially growing (unstable) mode. The same structure holds for the monopole, for which the critical angle is obtained in closed form whereas for the dipole it is computed numerically. Instabilities thus arise only if one imposes at the singularity a boundary condition sufficiently far from the Friedrichs one.

In this work, we  also applied  the same formalism to stealth BH solutions, where the metric is Schwarzschild despite a non-trivial scalar field profile. A notable example is the stealth solution with a non-constant  kinetic term $X$ \cite{Bakopoulos:2023fmv}, obtained within the same family of theories that yield BHs with primary hair. We  showed  that the dipolar perturbations of this stealth BH do not yield a propagating degree of freedom, which we explained through an ADM decomposition.

Interestingly, we found a similar conclusion for dipolar perturbations of the BH solution in 4D Einstein-Gauss-Bonnet theory \cite{Charmousis:2021npl} when applying our formalism; in this case, however, this property is specific to the BH solution and not to the theory, which suggests a strong coupling of the solution.

In addition, we revisited prior stealth solutions \cite{Minamitsuji:2019shy} identified in the literature, this time with a constant $X$. At the level of linear perturbations, the corresponding DHOST theories can be characterised by just three constant parameters. Except for special values of these parameters where no propagating mode exists, one generally finds a propagating degree of freedom. 
Similarly to GR, the Schr\"odinger radial coordinate now spans the entire real line but  it now covers the region outside an effective horizon that does not coincide with the horizon of the metric. This effective horizon depends on the scalar charge and the theory parameters. Again, we find that the mode is stable linearly.  

Our results on the dipolar stability of compact objects in DHOST complement our recent work on axial and radial perturbations around the same solutions \cite{Charmousis:2025xug,Iteanu2026rph}. To complete the linear analysis, it remains to  study the multipolar perturbations, which we leave for future work. However, with the definition of the master variable for both the monopole and the dipole, we are ready to tackle the problem.
The multipolar analysis will allow us to move from a $1+1$ dimensional problem to a $4D$ problem as well as to understand the coupling between the tensorial and the scalar degrees of freedom.

Several other directions are suggested by our results. 
First, since the function $n$, which defines the Schrödinger radial coordinate, grows as $r^2$ at large $r$ \eqref{r_star_hairy}, the effective scalar geometry has a timelike boundary at $r_*=\rmax$, as in anti-de Sitter spacetime, and the spectrum is purely discrete, with neither quasinormal modes nor late-time tails.
In anti-de Sitter spacetime, this structure, together with an equally spaced spectrum, leads to a weakly turbulent instability at the nonlinear level~\cite{Bizo__2011}. 
Whether a similar mechanism is at work here is an interesting open question, which would require going beyond the linear perturbations.
Second, neither the monopole nor the dipole propagates at linear order around the 4DEGB BH. Contrary to the $\nth=1/2$ theory, for which this can be explained by its ``extended cuscuton'' structure, we do not have a non-perturbative explanation for this property.
Going to higher order in perturbation theory would allow us to determine whether it reflects the absence of the scalar degree of freedom, or rather a strong coupling issue.
Third, it would also be interesting to study how the normal modes of the dipole and monopole are excited by astrophysical sources, for instance by a compact object orbiting the black hole, as in extreme mass-ratio inspirals (see \cite{Maselli_2020}). 
Finally, to complement this perturbative approach, we could also assess the stability of the solutions through a thermodynamical study.

\subsubsection*{Acknowledgments}
We would like to warmly thank Karim Noui, Christos Charmousis and David Langlois for the very helpful discussions.
We would also like to thank Jibril Ben Achour and Hugo Roussille for their interesting suggestions and advice.
We were supported by the ENS de Lyon postdoctoral fellowship during this work.

\begin{appendices}
\section{Component transformation rules}
\label{app:Component_transfo}
\noindent 
An infinitesimal even-parity coordinate transformation,
    $x^\mu \longrightarrow x^\mu+\xi^\mu,$
is parametrized, for each \((\ell,m)\), by three functions \(T(t,r)\), \(R(t,r)\), and \(\Theta(t,r)\), according to
\begin{equation}
   \xi^\mu=\left(T\,Y_{\ell m},\;R\,Y_{\ell m},\;\frac{\Theta}{r^2}\,\gamma^{\theta b}\partial_bY_{\ell m},\;\frac{\Theta}{r^2}\,\gamma^{\varphi b}\partial_bY_{\ell m}\right).
    \label{eq:gauge_vector_even}
\end{equation}
Under this transformation, the metric perturbation changes as
\begin{equation}
    h_{\mu\nu}
    \longrightarrow
    h_{\mu\nu}
    -\nabla_\mu\xi_\nu
    -\nabla_\nu\xi_\mu.
    \label{eq:metric_gauge_transformation}
\end{equation}
For the even-parity perturbation variables introduced in the main text, this gives
\begin{align}
H_0(t, r) &\to H_0(t, r)
+ 2\,\dot{T}(t, r)
+ \frac{\cA'}{\cA}\, R(t, r), \\[6pt]
H_1(t, r) &\to H_1(t, r)
+ \cA\, T^{\prime}(t, r)
- \frac{\dot{R}(t, r)}{\cB}, \\[6pt]
H_2(t, r) &\to H_2(t, r)
- \cA\,\frac{-R(t, r)\, \cB' + 2 \cB\, R^{\prime}(t, r)}{\cB^2}, \\[6pt]
K(t, r) &\to K(t, r)
- \frac{2}{r}\, R(t, r), \\[6pt]
G(t, r) &\to G(t, r)
- 2\, \Theta(t, r), \\[6pt]
\alpha(t, r) &\to \alpha(t, r)
- \frac{r R - 2 \cB\, \Theta + \cB\, r\, \Theta^{\prime}}{\cB\, r}, \\[6pt]
\beta(t, r) &\to \beta(t, r)
- \dot{\Theta}(t, r)
+ \cA\, T(t, r), \\[6pt]
\delta\phi(t, r) &\to \delta\phi(t, r)
- q\, T(t, r)
- R(t, r)\, \psi'(r).
\end{align}

On the unit two-sphere \((S^2,\gamma_{ab})\), the scalar spherical harmonics \(Y_{\ell m}\) satisfy
\begin{equation}
    \nabla^a\nabla_a Y_{\ell m}
    =
    -\ell(\ell+1)Y_{\ell m},
\end{equation}
where \(\nabla_a\) denotes the covariant derivative associated with the metric \(\gamma_{ab}\). In the particular case of the dipole, \(\ell=1\), one further has
\begin{equation}
    \nabla_a\nabla_b Y_{1m}
    =
    -\gamma_{ab}Y_{1m}.
    \label{eq:dipole_harmonic_identity}
\end{equation}
This identity is responsible for the degeneracy of the usual even-parity tensor harmonics in the dipolar sector, since the angular components of the metric perturbation can be written schematically as
\begin{equation}
    h_{ab}^{\ell=1}
    \propto
    \left(r^2 K-G\right)
    \gamma_{ab}Y_{1m}.
\end{equation}
Consequently, \(K\) and \(G\) do not correspond to two independent perturbation variables for \(\ell=1\): the angular metric perturbation depends only on the single combination $r^2 K-G$.

\section{Dipole equations and Monopole S-functions for quadratic Horndeski theories}
\label{app:S_Horndeski}
In this appendix, we give the explicit expressions, for a generic quadratic Horndeski theory with $X'\neq0$, of the fields introduced in Sec.~\ref{perturbations section}, together with the Schr\"odinger coordinates and the $S$-decomposition of the potentials of Sec.~\ref{sec:Schro_equation}, for both the monopole and the dipole.

\subsection{Field definitions}
The fields introduced in Sec.~\ref{perturbations section} read
\begin{eqnarray}
\label{app:nnvv}
\vv&=& \dot{\alpha}-\beta'+\frac{2 \cA \psi ' }{q r}\alpha+\frac{2  }{r}\beta -H_1\,,\\
    \zv&= &\vv-\frac{2 \cA \psi '  }{q r}\alpha-\frac{2 Z^2 }{\gamma ^2 r \cA}\beta +2 H_1\,, \\
\nv&=&\frac{ q^2 Z_X-\cA \left(2 X Z_X+Z\right)}{q Z_X \psi '}H_2-2 \cA H_1+\frac{2 Z^3 \cA  }{\gamma ^2 q r Z_X \psi '}\alpha+\frac{2 Z^2  }{\gamma ^2 r}\beta\, ,\\
    \mv&=& \frac{ Z_X \left(q^2-2 X \cA\right)}{2 Z \cA^2}\nv+\frac{\zv}{2}\,.
\end{eqnarray}
In terms of the original metric components, the master variable $\mv$ can be expressed as
\begin{equation}
    2\mv= \dot{\alpha}-\beta'
    +\frac{\gamma ^2\cA \psi ' \left(\gamma ^2 \cA Z_X \psi '^2-Z^3\right)}{q Z^5} H_2 
    + \left(1-\frac{2 \gamma ^2 \cA Z_X \psi '^2}{Z^3}\right)H_1
    +\frac{2   \cA \psi '}{q r}\alpha
    +\frac{2  }{r}\left(1-\frac{Z^2}{\gamma ^2 \cA}+\frac{Z_X \psi '^2}{Z}\right)\beta \, .
\end{equation}
Solving the equation $\mathscr{E}_3=0$ for $\zv$, one obtains the explicit relation \eqref{eq:zv_symbolicsolvesigma}
\begin{eqnarray}
    \zv&=& \frac{4 \gamma ^2 q r  \cA \psi '}{\gamma ^2 q^2 r \cA'+2 Z^2 \left(q^2-2 X \cA\right)}\dot{\mv}+\frac{2 \gamma ^2  q^2 r \cA'}{\gamma ^2 q^2 r \cA'+2 Z^2 \left(q^2-2 X \cA\right)}\mv\,.
\end{eqnarray}
\subsection{Schr\"odinger equation for the monopole and the dipole}
\label{app:Schro_horndeski}
We introduce the following notations which will prove useful in the multipole analysis:
\begin{equation}
\GT=\Ftwo-2X\Ftwo_X,\qquad \FT=\Ftwo,\qquad \mu_T=X\,\GTX,\qquad \nu_T=X^2\,\GTXX,\qquad m_F=X\FTX=\tfrac12(\FT-\GT),
\label{app:S_EFTfunctions}
\end{equation}
as well as the combinations
\begin{align}
\OmB&\equiv q^2\gamma^2-2X\GT^2\,, &
\mathcal{P}&\equiv X\frac{\mathrm d}{\mathrm dX}(\Ftwo\GT)\,, &
\kappa_T&\equiv \GT\,\nu_T-3\mu_T^2\,,
\label{app:S_comb1}\\
\EB&\equiv \cA\,(\GT+2\mu_T)-\frac{q^2\mu_T}{X}\,, &
\YB&\equiv 2X\GT\,\OmB+rX'q^2\gamma^2(\GT+2\mu_T)\,, &
\DB&\equiv(\GT+2\mu_T)\,\mathcal{P}\,\YB+2rX'\,\GT^2\,\OmB\,\kappa_T\,.
\label{app:S_comb2}
\end{align}
To pass to a Schr\"odinger equation \eqref{waveop} for a Horndeski theory with $X'\neq0$, we do the following change of coordinates:
\begin{align}
        W_{\ell=1}=W_{\ell=0}&=\frac{\psi'}{q}+\frac{\YB}{q\,\gamma\,\OmB\sqrt{q^2-2X\cA}}\,,\\[6pt]
n_{\ell=1}=n_{\ell=0}&=n=\sqrt{\frac{(q^2-2X\cA)\,\OmB^2\,\mathcal{P}}{r\,X'\,\DB}}\,,
\label{app:SametandNHorn}
\end{align}
Once in the right coordinate system, one can compute the effective potential $\Veff$ in terms of an $S$-function.
\begin{equation}
\Veff=S^2-\partial_*S+\Vs\,.
\label{app:S_def}
\end{equation}
We introduce the useful function
\begin{equation}
\Xi(r)=\frac{2r\,\mathcal{P}}{X\GT(\GT+2\mu_T)}
+2\int^r\mathrm{d}r\;r\,X'\,\frac{(q^2\gamma^2-X\GT^2)\,\mathcal{P}}{X^2\GT^2\,\OmB}\,.
\end{equation}
The integral depends on $r$ only through $X(r)$ and the factor $rX'$.
It is not local in general, as for the function $\Xi_p$ of the hairy black holes.
The $S$-function of the monopole is then given in a unique way by
\begin{equation}
\smono(r)=-\partial_*\ln\!\left[\frac{X\,\GT(\GT+2\mu_T)\,\sqrt{|\OmB/\DB|}\;\big|\Xi(r)+c_S\big|}{\sqrt{n}}\right],
\qquad \Vs=0\,,
\label{app:S_S0}
\end{equation}
where the constant $c_S$ is chosen such that $\Xi+c_S$ does not vanish on the domain
under consideration.

For the dipole, as seen from the examples treated in this article, one needs $\Vs\neq0$ and such a separation between the $S$ part and the $\Vs$ part is completely arbitrary. We still give an explicit compact formula which can be used to compute $\Veff$ for the dipole.
 We introduce
\begin{equation}
D_1\equiv 2\big(\gamma^2\EB-\GT^3\big)-r\gamma^2(\GT+2\mu_T)\,\cA'\,.
\label{app:S_D1}
\end{equation}
We choose
\begin{equation}
   \sdipo(r)= -\partial_*\ln\!\left[\frac{\sqrt{n\,|\DB|}}{r\,X\,\GT\,|D_1|}\right],
\label{app:S_S1}
\end{equation}
for which
\begin{equation}
\Vs=\frac{\mathcal{P}\,\YB\,\big(\gamma^2\EB-\GT^3\big)+\gamma^2r^2X'\,\GT^2\,\kappa_T\,\OmB\,\cA'}{\gamma^2\,r^2\,\DB}\,.
\label{app:S_Vs1}
\end{equation}
Here, we assume that on the domain under consideration, $n^2>0$, and $\OmB$, $\mathcal{P}$, $\DB$ do not vanish. Also in order for $\sdipo$ to be regular, one needs $D_1\neq0$ on the whole domain.
Positivity of the Schrödinger operator further requires $\Vs\geq0$ and vanishing boundary terms (behaviour of $\sdipo$ at the endpoints of the $r_*$ interval).
We insist that both have to be checked on each background.

One can check that for the Horndeski version of the $p=2$ black hole with primary hair described in Sec. 4.2.1 of Ref.~\cite{Charmousis:2025xug}, the dipole potential obtained from the master equation coincides with~\eqref{SandVsdef_Hairy}. Moreover, \eqref{app:S_S1} and \eqref{app:S_Vs1} reduce exactly to~\eqref{SandVsdef_Hairy}.

\end{appendices}

\newpage
\bibliographystyle{utphys}
\bibliography{References}

@article{Iyonaga_2018,
   title={Extended cuscuton: formulation},
   volume={2018},
   ISSN={1475-7516},
   url={http://dx.doi.org/10.1088/1475-7516/2018/12/002},
   DOI={10.1088/1475-7516/2018/12/002},
   number={12},
   journal={Journal of Cosmology and Astroparticle Physics},
   publisher={IOP Publishing},
   author={Iyonaga, Aya and Takahashi, Kazufumi and Kobayashi, Tsutomu},
   year={2018},
   month=Dec, pages={002–002} }

@article{Bakopoulos:2023fmv,
    author = "Bakopoulos, Athanasios and Charmousis, Christos and Kanti, Panagiota and Lecoeur, Nicolas and Nakas, Theodoros",
    title = "{Black holes with primary scalar hair}",
    eprint = "2310.11919",
    archivePrefix = "arXiv",
    primaryClass = "gr-qc",
    doi = "10.1103/PhysRevD.109.024032",
    journal = "Phys. Rev. D",
    volume = "109",
    number = "2",
    pages = "024032",
    year = "2024"
}

@misc{Iteanu2026rph,
      title={Radial Perturbations of Black Holes in DHOST Theories}, 
      author={Christos Charmousis and Simon Iteanu and David Langlois and Karim Noui},
      year={2026},
      eprint={2606.25972},
      archivePrefix={arXiv},
      primaryClass={gr-qc},
      url={https://arxiv.org/abs/2606.25972}, 
}

@book{Reed:1975mmp2,
    author = "Reed, Michael and Simon, Barry",
    title = "{Methods of Modern Mathematical Physics. {II}: Fourier Analysis, Self-Adjointness}",
    publisher = "Academic Press",
    address = "New York",
    year = "1975",
    isbn = "978-0-12-395826-6"
}

@article{Kimura:2017uor,
    author = "Kimura, Masashi",
    title = "{A simple test for stability of black hole by $S$-deformation}",
    eprint = "1706.01447",
    archivePrefix = "arXiv",
    primaryClass = "gr-qc",
    doi = "10.1088/1361-6382/aa903f",
    journal = "Class. Quant. Grav.",
    volume = "34",
    number = "23",
    pages = "235007",
    year = "2017"
}

@article{Bizo__2011,
   title={Weakly Turbulent Instability of Anti–de Sitter Spacetime},
   volume={107},
   ISSN={1079-7114},
   url={http://dx.doi.org/10.1103/PhysRevLett.107.031102},
   DOI={10.1103/physrevlett.107.031102},
   number={3},
   journal={Physical Review Letters},
   publisher={American Physical Society (APS)},
   author={Bizoń, Piotr and Rostworowski, Andrzej},
   year={2011},
   month=July }

@article{Motohashi_2016,
   title={Fundamental theorem on gauge fixing at the action level},
   volume={94},
   ISSN={2470-0029},
   url={http://dx.doi.org/10.1103/PhysRevD.94.124021},
   DOI={10.1103/physrevd.94.124021},
   number={12},
   journal={Physical Review D},
   publisher={American Physical Society (APS)},
   author={Motohashi, Hayato and Suyama, Teruaki and Takahashi, Kazufumi},
   year={2016},
   month=Dec }

@article{Kimura:2018eiv,
    author = "Kimura, Masashi and Tanaka, Takahiro",
    title = "{Robustness of the $S$-deformation method for black hole stability analysis}",
    eprint = "1805.08625",
    archivePrefix = "arXiv",
    primaryClass = "gr-qc",
    reportNumber = "KUNS-2723, YITP-18-52",
    doi = "10.1088/1361-6382/aadc13",
    journal = "Class. Quant. Grav.",
    volume = "35",
    number = "19",
    pages = "195008",
    year = "2018"
}

@article{Ishibashi:2003ap,
    author = "Ishibashi, Akihiro and Kodama, Hideo",
    title = "{Stability of higher dimensional Schwarzschild black holes}",
    eprint = "hep-th/0305185",
    archivePrefix = "arXiv",
    doi = "10.1143/PTP.110.901",
    journal = "Prog. Theor. Phys.",
    volume = "110",
    pages = "901--919",
    year = "2003"
}

@article{Baake:2023zsq,
    author = "Baake, Olaf and Cisterna, Adolfo and Hassaine, Mokhtar and Hernandez-Vera, Ulises",
    title = "{Endowing black holes with beyond-Horndeski primary hair: An exact solution framework for scalarizing in every dimension}",
    eprint = "2312.05207",
    archivePrefix = "arXiv",
    primaryClass = "hep-th",
    doi = "10.1103/PhysRevD.109.064024",
    journal = "Phys. Rev. D",
    volume = "109",
    number = "6",
    pages = "064024",
    year = "2024"
}

@article{DeFelice:2018ewo,
    author = "De Felice, Antonio and Langlois, David and Mukohyama, Shinji and Noui, Karim and Wang, Anzhong",
    title = "{Generalized instantaneous modes in higher-order scalar-tensor theories}",
    eprint = "1803.06241",
    archivePrefix = "arXiv",
    primaryClass = "hep-th",
    doi = "10.1103/PhysRevD.98.084024",
    journal = "Phys. Rev. D",
    volume = "98",
    number = "8",
    pages = "084024",
    year = "2018"
}

@article{Kobayashi:2014wsa,
    author = "Kobayashi, Tsutomu and Motohashi, Hayato and Suyama, Teruaki",
    title = "{Black hole perturbation in the most general scalar-tensor theory with second-order field equations II: the even-parity sector}",
    eprint = "1402.6740",
    archivePrefix = "arXiv",
    primaryClass = "gr-qc",
    reportNumber = "RESCEU-5-14, RUP-14-4, RESCEU-5/14",
    doi = "10.1103/PhysRevD.89.084042",
    journal = "Phys. Rev. D",
    volume = "89",
    number = "8",
    pages = "084042",
    year = "2014"
}

@article{Maselli_2020,
   title={Detecting Scalar Fields with Extreme Mass Ratio Inspirals},
   volume={125},
   ISSN={1079-7114},
   url={http://dx.doi.org/10.1103/PhysRevLett.125.141101},
   DOI={10.1103/physrevlett.125.141101},
   number={14},
   journal={Physical Review Letters},
   publisher={American Physical Society (APS)},
   author={Maselli, Andrea and Franchini, Nicola and Gualtieri, Leonardo and Sotiriou, Thomas P.},
   year={2020},
   month=Sept }

@article{Noui:2023ksf,
    author = "Noui, Karim and Roussille, Hugo and Langlois, David",
    title = "{Axial perturbations of black holes in scalar-tensor gravity: near-horizon behaviour}",
    eprint = "2304.03055",
    archivePrefix = "arXiv",
    primaryClass = "gr-qc",
    doi = "10.1088/1475-7516/2023/11/040",
    journal = "JCAP",
    volume = "11",
    pages = "040",
    year = "2023"
}

@article{Bakopoulos:2023sdm,
    author = "Bakopoulos, Athanasios and Chatzifotis, Nikos and Nakas, Theodoros",
    title = "{Compact objects with primary hair in shift and parity symmetric beyond Horndeski gravities}",
    eprint = "2312.17198",
    archivePrefix = "arXiv",
    primaryClass = "gr-qc",
    doi = "10.1103/PhysRevD.110.024044",
    journal = "Phys. Rev. D",
    volume = "110",
    number = "2",
    pages = "024044",
    year = "2024"
}

@article{Charmousis:2025xug,
    author = "Charmousis, Christos and Iteanu, Simon and Langlois, David and Noui, Karim",
    title = "{Axial perturbations of black holes with primary scalar hair}",
    eprint = "2503.22348",
    archivePrefix = "arXiv",
    primaryClass = "gr-qc",
    reportNumber = "CERN-TH-2025-058",
    doi = "10.1088/1475-7516/2025/05/102",
    journal = "JCAP",
    volume = "05",
    pages = "102",
    year = "2025"
}

@article{Takahashi:2021bml,
    author = "Takahashi, Kazufumi and Motohashi, Hayato",
    title = "{Black hole perturbations in DHOST theories: master variables, gradient instability, and strong coupling}",
    eprint = "2106.07128",
    archivePrefix = "arXiv",
    primaryClass = "gr-qc",
    reportNumber = "YITP-21-38",
    doi = "10.1088/1475-7516/2021/08/013",
    journal = "JCAP",
    volume = "08",
    pages = "013",
    year = "2021"
}

@article{Langlois:2015cwa,
    author = "Langlois, David and Noui, Karim",
    title = "{Degenerate higher derivative theories beyond Horndeski: evading the Ostrogradski instability}",
    eprint = "1510.06930",
    archivePrefix = "arXiv",
    primaryClass = "gr-qc",
    doi = "10.1088/1475-7516/2016/02/034",
    journal = "JCAP",
    volume = "02",
    pages = "034",
    year = "2016"
}

@article{Gleyzes:2014dya,
    author = "Gleyzes, J\'er\^ome and Langlois, David and Piazza, Federico and Vernizzi, Filippo",
    title = "{Healthy theories beyond Horndeski}",
    eprint = "1404.6495",
    archivePrefix = "arXiv",
    primaryClass = "hep-th",
    doi = "10.1103/PhysRevLett.114.211101",
    journal = "Phys. Rev. Lett.",
    volume = "114",
    number = "21",
    pages = "211101",
    year = "2015"
}

@article{Zumalac_rregui_2014,
   title={Transforming gravity: From derivative couplings to matter to second-order scalar-tensor theories beyond the Horndeski Lagrangian},
   volume={89},
   ISSN={1550-2368},
   url={http://dx.doi.org/10.1103/PhysRevD.89.064046},
   DOI={10.1103/physrevd.89.064046},
   number={6},
   journal={Physical Review D},
   publisher={American Physical Society (APS)},
   author={Zumalacárregui, Miguel and García-Bellido, Juan},
   year={2014},
   month=Mar }

@misc{bahamonde2026nonreciprocalblackholesprimary,
      title={Nonreciprocal black holes with primary hair in beyond Horndeski gravity: Exact solutions and stability}, 
      author={Sebastian Bahamonde and Theodoros Nakas and Simon Iteanu},
      year={2026},
      eprint={2610.11712},
      archivePrefix={arXiv},
      primaryClass={gr-qc},
      url={https://arxiv.org/abs/2610.11712}, 
}

@article{Gleyzes:2014qga,
    author = "Gleyzes, J\'er\^ome and Langlois, David and Piazza, Federico and Vernizzi, Filippo",
    title = "{Exploring gravitational theories beyond Horndeski}",
    eprint = "1408.1952",
    archivePrefix = "arXiv",
    primaryClass = "astro-ph.CO",
    doi = "10.1088/1475-7516/2015/02/018",
    journal = "JCAP",
    volume = "02",
    pages = "018",
    year = "2015"
}

@article{Horndeski:1974wa,
    author = "Horndeski, Gregory Walter",
    title = "{Second-order scalar-tensor field equations in a four-dimensional space}",
    doi = "10.1007/BF01807638",
    journal = "Int. J. Theor. Phys.",
    volume = "10",
    pages = "363--384",
    year = "1974"
}

@article{BenAchour:2016cay,
    author = "Ben Achour, Jibril and Langlois, David and Noui, Karim",
    title = "{Degenerate higher order scalar-tensor theories beyond Horndeski and disformal transformations}",
    eprint = "1602.08398",
    archivePrefix = "arXiv",
    primaryClass = "gr-qc",
    doi = "10.1103/PhysRevD.93.124005",
    journal = "Phys. Rev. D",
    volume = "93",
    number = "12",
    pages = "124005",
    year = "2016"
}

@article{Gibbons_2005,
   title={On the Stability of Naked Singularities with Negative Mass},
   volume={113},
   ISSN={1347-4081},
   url={http://dx.doi.org/10.1143/PTP.113.963},
   DOI={10.1143/ptp.113.963},
   number={5},
   journal={Progress of Theoretical Physics},
   publisher={Oxford University Press (OUP)},
   author={Gibbons, G. W. and Hartnoll, S. A. and Ishibashi, A.},
   year={2005},
   month=May, pages={963–978} }

@book{Teschl:2014,
    author    = "Teschl, Gerald",
    title     = "{Mathematical Methods in Quantum Mechanics: With Applications to Schr\"odinger Operators}",
    edition   = "2nd",
    series    = "Graduate Studies in Mathematics",
    volume    = "157",
    publisher = "American Mathematical Society",
    address   = "Providence, RI",
    year      = "2014"
}

@article{Niessen:1992,
  author  = {Niessen, Heinz-Dieter and Zettl, Anton},
  title   = {Singular {S}turm--{L}iouville problems: the {F}riedrichs extension and comparison of eigenvalues},
  journal = {Proceedings of the London Mathematical Society (3)},
  volume  = {64},
  number  = {3},
  pages   = {545--578},
  year    = {1992}
}

@article{Gesztesy:2019,
  author  = {Gesztesy, Fritz and Littlejohn, Lance L. and Nichols, Roger},
  title   = {On self-adjoint boundary conditions for singular {S}turm--{L}iouville operators bounded from below},
  journal = {Journal of Differential Equations},
  volume  = {269},
  number  = {9},
  pages   = {6448--6491},
  year    = {2020},
  eprint  = {1910.13117},
  archivePrefix = {arXiv},
  primaryClass  = {math.SP}
}

@misc{Iteanu2026Git,
    author = "Charmousis, Christos and Iteanu, Simon and Langlois, David and Noui, Karim",
    title = "{Radial Perturbations of Black Holes in Horndeski Theories}",
    howpublished = "GitHub",
    url = "https://github.com/SimonIteanuIncerti/Radial_perturbation_Horndeski",
    year = "2026",
    note = "Mathematica notebook for the computation of monopole perturbation equations and Schrodinger-like potentials in quadratic Horndeski theories"
}

@article{Charmousis:2021npl,
    author = "Charmousis, Christos and Leh\'ebel, Antoine and Smyrniotis, Evangelos and Stergioulas, Nikolaos",
    title = "{Astrophysical constraints on compact objects in 4D Einstein-Gauss-Bonnet gravity}",
    eprint = "2109.01149",
    archivePrefix = "arXiv",
    primaryClass = "gr-qc",
    doi = "10.1088/1475-7516/2022/02/033",
    journal = "JCAP",
    volume = "02",
    number = "02",
    pages = "033",
    year = "2022"
}

@article{Kobayashi_2011,
   title={Generalized G-Inflation: --Inflation with the Most General Second-Order Field Equations--},
   volume={126},
   ISSN={1347-4081},
   url={http://dx.doi.org/10.1143/PTP.126.511},
   DOI={10.1143/ptp.126.511},
   number={3},
   journal={Progress of Theoretical Physics},
   publisher={Oxford University Press (OUP)},
   author={Kobayashi, T. and Yamaguchi, M. and Yokoyama, J.},
   year={2011},
   month=Sept, pages={511–529} }

@article{Langlois:2015skt,
    author = "Langlois, David and Noui, Karim",
    title = "{Hamiltonian analysis of higher derivative scalar-tensor theories}",
    eprint = "1512.06820",
    archivePrefix = "arXiv",
    primaryClass = "gr-qc",
    doi = "10.1088/1475-7516/2016/07/016",
    journal = "JCAP",
    volume = "07",
    pages = "016",
    year = "2016"
}

@article{Minamitsuji:2019shy,
    author = "Minamitsuji, Masato and Edholm, James",
    title = "{Black hole solutions in shift-symmetric degenerate higher-order scalar-tensor theories}",
    eprint = "1907.02072",
    archivePrefix = "arXiv",
    primaryClass = "gr-qc",
    doi = "10.1103/PhysRevD.100.044053",
    journal = "Phys. Rev. D",
    volume = "100",
    number = "4",
    pages = "044053",
    year = "2019"
}

@article{Khoury:2020aya,
    author = "Khoury, Justin and Trodden, Mark and Wong, Sam S. C.",
    title = "{Existence and instability of hairy black holes in shift-symmetric Horndeski theories}",
    eprint = "2007.01320",
    archivePrefix = "arXiv",
    primaryClass = "astro-ph.CO",
    doi = "10.1088/1475-7516/2020/11/044",
    journal = "JCAP",
    volume = "11",
    pages = "044",
    year = "2020"
}

@article{deRham:2019gha,
    author = "de Rham, Claudia and Zhang, Jun",
    title = "{Perturbations of stealth black holes in degenerate higher-order scalar-tensor theories}",
    eprint = "1907.00699",
    archivePrefix = "arXiv",
    primaryClass = "hep-th",
    reportNumber = "Imperial/TP/2019/CdR/03",
    doi = "10.1103/PhysRevD.100.124023",
    journal = "Phys. Rev. D",
    volume = "100",
    number = "12",
    pages = "124023",
    year = "2019"
}

@article{Kral_Zettl,
author = {A. M. Krall and A. Zettl},
title = {{Singular selfadjoint Sturm-Liouville problems}},
volume = {1},
journal = {Differential and Integral Equations},
number = {4},
publisher = {Khayyam Publishing, Inc.},
pages = {423 -- 432},
year = {1988},
doi = {10.57262/die/1372451946},
URL = {https://doi.org/10.57262/die/1372451946}
}

@misc{Sirera:2024ghv,
    author = "Sirera, Sergi and Noller, Johannes",
    title = "{Stability and quasinormal modes for black holes with time-dependent scalar hair}",
    eprint = "2408.01720",
    archivePrefix = "arXiv",
    primaryClass = "gr-qc",
    month = "8",
    year = "2024"
}

\end{document}